\documentclass[reprint,superscriptaddress,nofootinbib,amsmath,amssymb,aps,prd]{revtex4-1}

\usepackage[T1]{fontenc}
\usepackage{lipsum}
\usepackage{enumitem}
\usepackage{ulem}
\usepackage{mathrsfs}
\usepackage{bm, bbm, latexsym, amsthm, tensor}
\usepackage{amsthm, thmtools}
\usepackage{fontspec}
\usepackage{graphicx}
\usepackage{float}
\usepackage{caption}
\usepackage[dvipsnames]{xcolor}
\usepackage{booktabs,empheq}
\usepackage{microtype}
\usepackage{slashed}
\usepackage{cancel}

\usepackage{hyperref}
\hypersetup{
    colorlinks,
    citecolor=cyan,
    linkcolor=blue,
    filecolor=magenta,      
    urlcolor=cyan,
    breaklinks=true,
}
\usepackage[capitalise]{cleveref}

\providecommand{\adsurl}[1]{}

\newcommand{\CD}{{\mathcal D}}
\newcommand{\CR}{{\mathcal R}}
\newcommand{\CW}{{\mathcal W}}
\newcommand{\CQ}{{\mathcal Q}}

\newcommand{\average}[1]{\left\langle #1 \right\rangle_\CD}

\newcommand{\initial}[1]{{#1_{\rm \bf i}}}

\newcommand{\Di}{{\CD_i}}
\newcommand{\DT}{{\CD_0}}
\newcommand{\av}[1]{\left\langle #1 \right\rangle_\CD}
\newcommand{\at}{\tilde a}
\newcommand{\rhoP}{\varrho^\CD_\Phi}
\newcommand{\pP}{p^\CD_\Phi}

\newcommand{\rd}{{\mathrm d}}

 1

\newcommand{\beq}{\begin{equation}}
\newcommand{\eeq}{\end{equation}}
\newcommand{\beqNo}{\begin{equation*}}
\newcommand{\eeqNo}{\end{equation*}}

\newcommand{\CK}{\mathcal K}

\newcommand{\I}{{\rm I}}
\newcommand{\II}{{\rm II}}

\usepackage{tensor,bm}

\usepackage{microtype}

\usepackage{hyperref}
\hypersetup{
    colorlinks,
    citecolor=NavyBlue,
    linkcolor=darkgreen,
    filecolor=magenta,      
    urlcolor=NavyBlue,
    breaklinks=true,
}
\usepackage[capitalise]{cleveref}

\begin{document}
\title{Closure of averaged Einstein equations by DESI data}

\author{Thomas Buchert}
\email{buchert@cosmunix.de}
\affiliation{Universit\'e Lyon 1, ENS de Lyon, CNRS, CRAL, UMR 5574, Lyon, France}

\definecolor{myblue}{rgb}{0.2,0.3,0.7}
\definecolor{darkgreen}{rgb}{0,0.3,0}
\definecolor{mygreen}{rgb}{0,0.5,0}
\definecolor{grey}{rgb}{0.5,0.5,0.5}
\definecolor{orange}{rgb}{1,0.5,0}
\definecolor{mybloodyred}{rgb}{0.7,0,0}
\definecolor{chromeyellow}{rgb}{1.0, 0.65, 0.0}
\definecolor{yellow(ncs)}{rgb}{1.0, 0.83, 0.0}
\definecolor{saffron}{rgb}{0.96, 0.77, 0.19}
\definecolor{laserlemon}{rgb}{1.0, 1.0, 0.13}
\newcommand{\thomas}[1]{{\color{mybloodyred} {[{\bf Thomas}: #1]} }}
\newcommand{\edt}[1]{{\color{mybloodyred}{#1}}}
%

\newcommand{\pn}{\nabla}
\newcommand{\tr}{\mathrm Tr}
\newcommand\eprintarXiv[1]{\href{http://arXiv.org/abs/#1}{arXiv:#1}}
\begin{abstract}
We benchmark the hypothesis that the redshift dependence of postulated dark energy in the standard model, measured by the DESI collaboration, is a natural outcome of dark energy-free models of structure formation in general relativity. 
We recall fundamental results and the properties of averaged dust cosmologies that suggest to replace the dark energy equation of state through the nonperturbative impact of the inhomogeneous geometry. The relevant mechanism can be phrased in terms of an effective mean field description of inhomogeneities using the fluid analogy of a scalar field (the `morphon'). Without violating energy conditions this effective field exhibits a cross-over to phantom behaviour, compatible with observations. The exact dark energy-free framework needs a single closure condition that we implement in terms of parametrizations of the 3-years data of the DESI team. The resulting redshift dependence of effective cosmological characteristics is then illustrated and subject to physical interpretation. The aim of this paper is to set a benchmark that has to be met by any structure formation model or simulation investigated within general relativity in a flow-orthogonal foliation of spacetime, given the hypothesis.
\end{abstract}
\maketitle
%
%

\section{Introduction}
\label{I}

Let us start by recalling the key results concerning the modeling of inhomogeneities within general relativity that led to the understanding of an emerging repulsive effect on the global properties of world models both from fundamental insights and explicit modeling. 

The first key result is that spatially averaged cosmologies do not result in a FLRW (Friedmann-Lema\^\i tre-Robertson-Walker) model that forms the basis of the standard $\Lambda$CDM model (cold dark matter with a cosmological constant $\Lambda$) of cosmology. Additional effective terms, dubbed cosmological backreaction terms, arise due to fluctuations in the expansion rate, fluctuations in shear and, most importantly, deviations from a constant Friedmannian curvature. (We here confine ourselves to an irrotational dust matter model \cite{buchert:dust} admitting a flow-orthogonal foliation of spacetime.\footnote{This framework covers all the physics here; for more general effective equations see \cite{buchert:fluid} by including pressure in a flow-orthogonal foliation, \cite{BMR} for general energy-momentum sources and a tilted $4$-velocity; for the covariance of the equations \cite{covariance}, and for the dependence on the choice of foliation \cite{foliations,mourierheinesen}.})
These backreaction terms globally vanish in Newtonian cosmology due to their property of being divergences on a flat space and the implementation of periodic boundary conditions on deviations from a background cosmological model \cite{buchertehlers}. This explains the absence of global effects also in many attempts using quasi-Newtonian weak field approximations or numerical simulations with 3-torus architecture due to the requirement that dynamical variables average out on an assumed FLRW background. A global effect has a genuinely general-relativistic, background-free and non-perturbative physical foundation.

Second, average spatial curvature emerges due to structure formation, even if initially it is set to zero \cite{buchert:dust,buchert:darkenergyEOS,buchert2008}, \cite{emergingcurvature}.  
For a universe model that is volume-dominated by voids today, the average spatial curvature on large scales is negative, for reviews see \cite{buchert2008}, \cite{buchert2011}, \cite{buchertrasanen}. In other words, the average scalar curvature does not obey a conservation law as in Friedmannian cosmology \cite{buchertcarfora2008}. It generically couples to the kinematics of structure formation. 

Third, the averaged effective cosmological equations admit two classes of exact scaling solutions. One class is non-generic; it corresponds to the curvature behavior of the standard model decoupled from the averaged inhomogeneities. The second class is generic where the averaged curvature `talks' to structure formation.
Moreover, the first class where the average features Friedmannian behavior forms a repellor in the space of average scaling models as a result of a dynamical systems analysis \cite{roy:instability}: the average is driven away towards attractors that lie in the region of globally isotropic (subdominant shear) and negative average curvature (dominant expansion rate fluctuations). We may call this the `dark energy sector' \cite{morphon}, \cite{buchert2008}, while on small scales, a dominant shear and positive average curvature results in a further attractor (the `dark matter sector' \cite{vigneronbuchert}). 
These scaling solutions can be employed to support dark energy-free fits to supernova observations \cite{desgrange}, and to explain the curvature inconsistencies if the Friedmannian curvature is allowed as a fitting parameter, and the Hubble tension \cite{heinesenbuchert}. However, scaling solutions lead to effective equations of state that are independent of redshift. Our closure strategy will result in a nonscaling solution. 

Fourth, a complementary description of nonperturbative geometric backreaction  can be exploited in terms of an effective scalar field (the morphon field) \cite{morphon} that may reveal a cross-over to phantom behavior---without there being any dark energy phantom field. We will also represent our findings through the morphon field.\\

This paper is organized as follows. Section \ref{II} recalls the averaged Einstein equations for an irrotational dust matter model and provides the framework to study the representation of the dark energy equation of state in terms of backreaction variables as well as the morphon field. Closure of the averaged system is then achieved through 
the 3-years DESI data using first as a showcase the Chevallier-Polarski-Linder (CPL) parametrization in Section \ref{III}. Section \ref{IV} then 
gives a list of results with an eye on the physical interpretation and puts results from structure formation models into perspective. Appendices are devoted to the robustness with respect to observational uncertainties (Appendix~\ref{uncertainties}), different parametrizations of the data (Appendix~\ref{robustness}), as well as a  nonparametric reconstruction of the results (Appendix~\ref{nonparametric}), a comparison with model-independent bounds  (Appendix~\ref{koksbang}), and a comment on the approximations entering the dynamical closure  (Appendix~\ref{distances}).

\section{Effective equations for inhomogeneous universe models}
\label{II}

We employ a foliation of spacetime, containing an irrotational fluid
of dust matter, into flow--orthogonal hypersurfaces with the 3-metric components
$h_{ij}$ in the line-element,
\begin{equation}
{}^4 {\bf g} = -\mathrm{d}t^2 + {^3}{\bf h}\;\;\;;\;\;\;{}^3 {\bf h} = h_{ij}\,\mathrm{d}X^i \otimes \mathrm{d}X^j\ .
\end{equation} 
We define the following \textit{volume averager}, restricting 
attention to scalar functions $\Psi (t,X^i )$:
\begin{equation}
\label{eq:average-GR}
\average{\Psi (t, X^i)}:=
\frac{1}{V_\CD}\int_\CD \Psi (t, X^i)\;J \mathrm{d}^3 X  \;\;;\;\;V_\CD = \int_\CD J \mathrm{d}^3 X \ ,
\end{equation}
with  $J:=\sqrt{\det(h_{ij})}$; $X^i$ are  coordinates that  are constant
along  flow lines, which are here spacetime geodesics. Averaging is
over a collection of free-falling fluid elements contained within 
a simply-connected compact domain $\CD$ in a spatial hypersurface.
Defining a \textit{volume scale factor} by the volume $V_\CD$ of such a
restmass preserving domain, normalized by the  volume of the initial domain,
$\initial\CD$,
\begin{equation}
\label{scalefactor}
a_\CD := \left(\frac{V_\CD}{V_{\initial\CD}}\right)^{1/3}\;\;,
\end{equation}
the following exact equations can be derived \cite{buchert:dust}
(an overdot denotes partial time-derivative). First, by averaging 
\textit{Raychaudhuri's equation} we obtain: 
\begin{equation}
\label{averageraychaudhuri}
3\frac{{\ddot a}_\CD}{a_\CD} + 4\pi G \langle\varrho\rangle_\CD - \Lambda
= {\cal Q}_\CD\;\;\;;\;\;
 \langle\varrho\rangle_\CD = \frac{M_\CD}{V_{\initial\CD}a_\CD^3}\ .
\end{equation}
The first integral of the above equation is directly given by averaging 
the \textit{energy constraint}:
\begin{equation}
\label{averagehamilton}
\left( \frac{{\dot a}_\CD}{a_\CD}\right)^2 - \frac{8\pi G}{3}\langle\varrho\rangle_\CD
- \frac{\Lambda}{3}=- \frac{\average{\CR} +{\cal Q}_\CD}{6} \ ,
\end{equation}
where the total restmass $M_\CD$, the averaged spatial 3-Ricci scalar
$\average{\CR}$   and   the  \textit{kinematical backreaction  term}  ${\cal Q}_\CD$  
are domain-depen\-dent and, except the mass, time-depen\-dent functions.
The backreaction source term reads:
\begin{equation}
\label{eq:Q-GR} 
{\cal Q}_\CD : = 2 \average{\II} - \frac{2}{3}\average{\I}^2 =
\frac{2}{3}\average{\left(\theta - \average{\theta}\right)^2 } - 
2\average{\sigma^2}\ ,
\end{equation}
where $\I = \Theta^i_{\;i}$  and $\II = \frac{1}{2}[\,(\Theta^i_{\;i})^2 - 
\Theta^i_{\;j}\Theta^j_{\;i}\,]$  
denote  the  principal scalar invariants  of the  expansion
tensor, defined  as minus the extrinsic
curvature $K_{ij}:=-\Theta_{ij}$. In the second equality of \eqref{eq:Q-GR}, $\CQ_\CD$  was split 
into kinematical invariants through the decomposition  
$\Theta_{ij} = \frac{1}{3}g_{ij}\theta + \sigma_{ij}$, with the rate of expansion 
$\theta =\Theta^i_{\;i}$, the shear tensor components $\sigma_{ij}$, and the rate of shear
$\sigma^2 :=1/2 \sigma_{ij}\sigma^{ij}$;
note that vorticity is absent in the present foliation; 
we adopt the summation convention.

The time-derivative of the averaged energy constraint (\ref{averagehamilton})
agrees with the averaged Raychaudhuri equation (\ref{averageraychaudhuri}) by virtue
of the following {\it integrability  condition}:
\begin{equation}
\label{integrability}
\frac{1}{a_\CD^6}\partial_t (a_\CD^6 {\CQ}_\CD) + \frac{1}{a_\CD^2}  
\partial_t (a_\CD^2 \average{\CR}) = 0 \ ,
\end{equation}
where we have introduced a \textit{volume Hubble functional} $H_{\cal D}:= 
{\dot a}_\CD / a_\CD$.
The above equations can formally be recast into standard
Friedmann equations for an effective perfect fluid energy momentum tensor
with effective sources \cite{buchert:fluid,buchert2008}:\footnote{In this
representation
of the effective equations ${p}_{\rm eff}$  denotes
an `effective (geometric) pressure'; there is no pressure due to a matter source here.}
\begin{subequations}
\label{equationofstate}
\begin{align}
\varrho^{\CD}_{\rm eff} &= \average{\varrho}-\frac{1}{16\pi G}{\CQ}_\CD - 
\frac{1}{16\pi G}{\cal W}_\CD \ , \\
{p}^{\CD}_{\rm eff} &=  -\frac{1}{16\pi G}{\CQ}_\CD + \frac{1}{48\pi
G}{\cal W}_\CD \ ;
\end{align}
\end{subequations}
\begin{subequations}
\label{effectivefriedmann}
\begin{align}
&3\frac{{\ddot a}_\CD}{a_\CD} = \Lambda - 4\pi G (\varrho^{\CD}_{\rm eff}
+3{p}^{\CD}_{\rm eff})\ ,\\
&3H_\CD^2 = -3\frac{k_{\initial\CD}}{a_\CD^2 } +\Lambda + 8\pi G \varrho^{\CD}_{\rm eff} \ ,\\
&{\dot\varrho}^{\CD}_{\rm eff} + 
3H_\CD \left(\varrho^{\CD}_{\rm eff}
+{p}^{\CD}_{\rm eff} \right)=0\ ,
\end{align}
\end{subequations}
with the intrinsic curvature deviation from constant-curvature ${\cal W}_\CD : = \average{\CR} - 6 k_{\initial\CD} / a_\CD^2$.
Eqs.~(\ref{effectivefriedmann}) correspond to the equations
(\ref{averageraychaudhuri}),
(\ref{averagehamilton}) and (\ref{integrability}), respectively. 

Given an equation of state of the form 
$p^\CD_{\rm eff} = \beta (\varrho^\CD_{\rm eff}, a_{\cal D})$
that relates the effective sources (\ref{equationofstate}) with a possible explicit
dependence on the volume scale factor, 
the effective system of equations (\ref{effectivefriedmann})
is closed (one of the equations (\ref{effectivefriedmann}) is redundant). This is the freedom we will spend on
DESI.

\subsection{The morphon picture}
\label{morphonic}

The backreaction terms that add to the fundamental sources in \eqref{equationofstate} suggest themselves to be written in terms of the fluid analogy to a minimally coupled scalar field,
dubbed \textit{morphon} $\Phi_\CD$ \cite{morphon}:
\begin{subequations}
	\label{morphon}
\begin{align}
\varrho^{\CD}_{\rm eff} &= \average{\varrho} +  \varrho^\Phi_\CD \ ; \ \varrho^\Phi_\CD :=\epsilon \frac{1}{2}\dot{\Phi}_\CD^2 + U_\CD\left(\Phi_\CD\right) \ , \\
{p}^{\CD}_{\rm eff} &= p^\Phi_\CD \ ; \ p^\Phi_\CD :=\epsilon \frac{1}{2}\dot{\Phi}_\CD^2 - U_\CD\left(\Phi_\CD\right)  \ ,
\end{align}
\end{subequations}
where $\epsilon = +1$ for a standard scalar field and $\epsilon = -1$ for a phantom field (with a negative kinetic energy density). This leads to the correspondence:\footnote{The change of variables $(\CW_\CD, \CQ_\CD) \rightarrow (\Phi_{\CD}, U_\CD)$ is not necessarily a diffeomorphism; there are cases where we have to change the morphon from a real to a phantom field, respectively from a phantom to a real field, each time $\CW_\CD + 3\CQ_\CD = -24\pi G \epsilon \dot\Phi_\CD^2$ changes its sign from negative to positive, respectively from positive to negative.}
\begin{align}
\label{morphoncorrespondence}
	\epsilon\dot{\Phi}^2_\CD = -\frac{\CW_\CD + 3\CQ_\CD}{24\pi G}  \ , \ 
	U_\mathcal{D} = \frac{-\CW_\CD}{24\pi G} \ . 
\end{align}
In the framework of the morphon field, the integrability condition \eqref{integrability}
becomes the Klein-Gordon equation:
\begin{align}
&\ddot{\Phi}_\CD + 3H_\CD \dot{\Phi}_\CD + \epsilon \frac{\rd U_\CD (\Phi_\CD)}{\rd \Phi_\CD} = 0\; .
\end{align}

\subsection{Reference to FLRW modeling}

We will replace a postulated dark energy component in the FLRW model with an emerging repulsive term in the averaged inhomogeneous cosmology framework without a cosmological constant. 
We hence compare the energy budget in the FLRW model 
with the domain-dependent energy budget in the inhomogeneous models:
\begin{align}
\label{omega}
&\Omega_m^{\CD}\;+\;\Omega_{\Lambda}^{\CD}\;+\;\Omega_k^\CD \;+\;\Omega_{\CW}^{\CD}\;+\;
\Omega_{\CQ}^{\CD}\;=\;1\;\;,\quad {\rm with:}\nonumber\\
&\Omega_m^{\CD} : = \frac{8\pi G  \langle\varrho\rangle_{\cal D} }{3 H_{\CD}^2}  \ ,\ 
\Omega_{\Lambda}^{\CD} := \frac{\Lambda}{3 H_{\CD}^2 }\ , \ \Omega_{k}^{\CD} := - \frac{k_{\initial\CD}}{a_\CD^2 H_{\CD}^2 }\ ,\nonumber\\ 
&\Omega_{\CW}^{\CD} := - \frac{\CW_\CD}{6 H_{\CD}^2 }\ , \
\Omega_{\CQ}^{\CD} := - \frac{{\CQ}_{\CD}}{6 H_{\CD}^2 } \ ,
\end{align}
where the first three arise in the FLRW model, independently of domain.
For the commonly adopted assumption of vanishing constant-curvature parameter ($k$ in the FLRW model) and our assumption of vanishing cosmological constant in the inhomogeneous models, we thus look at the correspondent expression to the dark energy content of the FLRW model:\footnote{Notice that dark energy corresponds to a positive sum of the adimensional averaged curvature deviation and kinematical backreaction functionals. Beware that these latter are defined with opposite signs to the bare variables. We expect  $\Omega_{\CW}^{\CD}>0$ and $\Omega_{\CQ}^{\CD} < 0$ in the dark energy sector.}
\beq
\label{gamma}
\Omega_{\CW}^{\CD}\;+\;\Omega_{\CQ}^{\CD} = : \Gamma^\CD_{\CW\CQ}\ , \ \gamma^\CD_{\CW\CQ}:=
\frac{\Gamma^\CD_{\CW\CQ}}{\Omega_m^{\CD}} \ ,
\eeq
where the latter is normalized to the matter density functional. 
It is related to the Friedmannian dark energy equation of state $w^F : = p^F_\Lambda / \varrho^F_\Lambda = -\Lambda / \Lambda = -1$. In our model, the effective equation of state (or morphon equation of state) is written in terms of physical variables, see Sect.~2.4.3 of \cite{buchert2008}, and \cite{buchert:darkenergyEOS}:
\beq
\label{wPhi}
w^{\Phi}_\CD : = \frac{\CQ_\CD - (1/3) \CW_\CD}{\CQ_\CD + \CW_\CD} = \frac{p^\Phi_\CD}{\varrho^\Phi_\CD}\ .
\eeq
This effective equation of state is nonexistent in homogeneous models.

The system of equations~\eqref{effectivefriedmann}, written in terms of a morphon field, is formally equivalent to the Friedmann equations sourced by dust matter and a postulated \textit{fundamental} and domain-independent homogeneous scalar field $\psi$ minimally coupled to a potential $U\left(\psi\right)$, \cite{madsen1987}, \cite{copeland:scaling}, \cite{sahni2000}: 
\begin{subequations}
	\label{eq::Fried_psi}
	\begin{align}
		\frac{\ddot{a}}{a} &= -\frac{8\pi G}{3} \left(\frac{\varrho_{\rm m}}{2} + \epsilon\dot{\psi}^2 - U(\psi) \right) \, ; \\
		&\ddot{\psi} + 3H\dot{\psi} + \epsilon\frac{\rd U(\psi)}{\rd \psi} = 0 \ ,
	\end{align}
with the energy constraint,
\begin{align}
 		H^2 = \frac{8\pi G}{3}\left(\varrho_{\rm m} + \epsilon \frac{1}{2}\dot{\psi}^2 + U(\psi) \right) \ ,
\end{align}
\end{subequations}
where $\varrho_{\rm m}$ is the matter density of the FLRW model. This system has been studied for the description of so-called quintessence models or for scenarii of inflation through specification of the form of the potential $U(\psi)$. In the effective models, the morphon potential is physically determined and is represented by averaged curvature \eqref{morphoncorrespondence}.

In the following we end the theory section by recalling closure of the system of equations \eqref{effectivefriedmann} through the generic class of exact scaling solutions.

\subsection{The space of scaling solutions}

Rewriting the integrability condition \eqref{integrability},
\begin{equation}
a_\CD^{-6}\partial_{t}\left(a_\CD^{6}{\CQ}_\CD\right)+
a_\CD^{-2}\partial_{t}\left(a_\CD^{2}\CW_\CD\right)\,=\,0\ ,
\end{equation}
where we replaced $\av{\CR}$ by $\CW_\CD$, 
a first scaling solution of that equation is obviously provided by:
\begin{equation}
\label{ndiffpsolution}
\CQ_\CD={\CQ}_{\initial\CD}\; a_\CD^{-6}\quad , \quad
\CW_\CD={\CW}_{\initial\CD}\; a_\CD^{-2}\ .
\end{equation} 
This degenerate case provides the only scaling solution that strictly decouples the kinematics of structure formation  from the averaged scalar curvature deviation, and it corresponds to the Friedmannian description of structure formation on a FLRW background with constant curvature.\footnote{We emphasize that, in the present framework as opposed to standard perturbation theory, this solution is background-free and non-perturbative, which implies that, through the volume scale factor $a_\CD$, averaged curvature and backreaction 
are still implicitly coupled by virtue of the effective equation
(\ref{averagehamilton}) that contains both terms as sources.}

On the other hand, in \cite{morphon}, we have studied
a class of solutions that directly couples kinematical backreaction and averaged
curvature functionals, prescribed in the form of scaling laws of the volume scale factor $a_{\CD}$.
This generic case mirrors the very essence of general relativity, namely the coupling of matter to curvature.  
Also, we restrict attention to the case $\Lambda = 0$, and we direct the reader to \cite{morphon}, \cite{buchert2008} and \cite{vigneronbuchert} for a full classification of the scaling solutions.\footnote{The paper \cite{vigneronbuchert} provides a typo-corrected version of the morphon potential in \cite{morphon}.}

The averaged dust matter density 
$\langle\varrho\rangle_\CD$ evolves, for a restmass preserving domain $\CD$, as 
$\langle\varrho\rangle_\CD=\langle\varrho\rangle_{\initial\CD}\; a_\CD^{-3}$. 
Suppose that the backreaction term and the averaged curvature deviation also obey scaling laws, 
\begin{equation}
\label{prescription}
{\CQ}_\CD={\CQ}_{\initial \CD}\; a_\CD^{n}\;\;\;\;;\;\;\;\;
{\CW}_\CD={\CW}_{\initial \CD}\; a_{\CD}^{p}\;\;,
\end{equation}
where ${\CQ}_{\initial\CD}$ and ${\CW}_{\initial\CD}$ 
denote the initial values of ${\CQ}_\CD$ and ${\CW}_\CD$, respectively.

The solution (\ref{ndiffpsolution}) is the only solution with $n\neq p$. Generic solutions obey
$n=p$. We define a coupling parameter $r_\CD$ (that can be chosen
differently for a chosen domain of averaging\footnote{For notational ease we henceforth
drop the index $\CD$ and simply write $r$.}) such that ${\CQ}_{\initial\CD}\propto
{\CW}_{\initial\CD}$; the solution reads: 
\begin{equation}
\label{backreactionparameter}
{\CQ}_\CD \;=\;r \;{\CW}_\CD\;=\;r \;{\CW}_{\initial\CD} \,a_\CD^{n}\ , 
\end{equation}
where
\begin{equation}
\label{neqpsolution}
n\;=\;-2 \frac{(1+3r)}{(1+r)}\;\ne -6 \ ,\quad r = - \frac{n+2}{n+6}\; \neq-1\ .
\end{equation}
We infer that the effective equation of state \eqref{wPhi} for the scaling solutions is constant, 
\beq
\label{EOSscaling}
w^\Phi_\CD = \frac{r-\frac{1}{3}}{r+1} = -\frac{1}{3} (n+3) \ .
\eeq
The case $n=0$ is equivalent to a scale-dependent cosmological constant; we have the relations $\CW_\CD = -3\Lambda_\CD$ and $\CQ_\CD = \Lambda_\CD$. The case $n=-2$ corresponds to the
evolution of a scale-dependent constant curvature model.

We refer the reader to Sect.~4 of \cite{vigneronbuchert} for the explicit forms of the morphon and its potential. For the modeling of present-day dark energy the relevant case of positive $\gamma^\CD_{\CW\CQ}$ (\textit{cf.} \eqref{gamma}), 
along with the condition $\epsilon = +1$ for $n < 0$, could be applied in the lower redshift regime where dark energy dominates. 
The reconstruction from DESI data will reveal a non-scaling behavior with a phantom crossing at some redshift.
The phantom branch is characterized by $\gamma^\CD_{\CW\CQ} > 0$, $n > 0$ and $\epsilon = -1$.

\section{Closure with DESI data}
\label{III}

We close the system of equations (\ref{effectivefriedmann})
using the DESI 3-years data \cite{DESI} in their basic parametrization. We evaluate the uncertainties in Appendix~\ref{uncertainties}, and Appendix~\ref{robustness} analyses the dependence on different parametrizations. 
Even a nonparametric reconstruction is possible that we outline in Appendix~\ref{nonparametric}. 
Appendix~\ref{koksbang} gives a closure-prediction of a bound that is obtained from a model-independent method, and Appendix~\ref{distances} extends the analysis of the only approximation used for the closure. This appended material shows that we can read the following showcase parametrization as being benchmark-representative with no substantial changes of the main results, and it confirms the DESI team's statement that their base parametrization is representative if compared against other parametrizations.

\subsection{Showcase DESI parametrization}

The simplest (and DESI's baseline) parametrization of a dynamical dark-energy equation
of state is Chevallier--Polarski--Linder \cite{CP,CPL},
\begin{subequations}
\begin{align}
 w_{\rm DESI}(a)&=w_0+w_a\,(1-a) \; ;\\
 w_{\rm DESI}(z)&=w_0+w_a\,\frac{z}{1+z}\quad ,\quad  a=\frac{1}{1+z}\; ,
 \label{eq:cpl}
\end{align}
with $a$ the Friedmannian scale factor normalized to $a(z{=}0)=1$. The
corresponding density factor follows from
$\dot\varrho_{\rm DE}+3H(1+w)\varrho_{\rm DE}=0$:
\begin{align}
 f(a)\;:=\;\frac{\varrho_{\rm DE}(a)}{\varrho_{\rm DE}(1)}
 &=\;\exp\!\left[-3\!\int_1^a\!\frac{1+w_{\rm DESI}(a')}{a'}\,\mathrm{d}a'\right] \nonumber \\
 &=\; a^{-3(1+w_0+w_a)}\,e^{-3w_a(1-a)} \; .
 \label{eq:f}
\end{align}
\end{subequations}

\subsection{Approximations underlying the single closure condition}
\label{onlyapprox}

DESI parametrizations are built on the Friedmannian scale factor, normalized to $a(t_0)=1$ at $z=0$, while the system  (\ref{effectivefriedmann}) involves the volume scale factor $a_\CD$ that is normalized to $a_\CD (t_{\rm i})= 1$ at some initial time $t_{\rm i}$. Furthermore, while $a(t)$ is the same in all spacetime directions due to the high symmetry of the Friedmann models, $a_\CD$ is spatial. To bridge the gap we will assume, following R\"as\"anen \cite{rasanen2009,rasanen2010}:
\begin{equation}
1+z=\frac{a_\DT}{a_\CD}
 \quad\Longleftrightarrow\quad
 \at\;:=\;\frac{a_\CD}{a_\DT}=\;\frac{1}{1+z}\;=\;a \; .
 \label{eq:rasanen}
\end{equation}
This is the only approximation entering the \textit{dynamical} closure. We will, however, also perform fits 
that rest on transverse distances, which brings in R\"as\"anen's companion result for
angular diameter distances. This needs in addition that the null shear is negligible (see the list of caveats below). Appendix~\ref{distances} quantifies this, and proposes closure fits that remove the dependence altogether.

Furthermore, \eqref{eq:rasanen} is the only place where the different normalizations enter, 
hence it is not $a_\CD$ but the ratio $\at := a_\CD / a_\DT$ that may be substituted for $a$
in \eqref{eq:cpl} and \eqref{eq:f} (and in all previous equations if interpreted with this normalization): 
\begin{equation}
a_\DT = \left(\frac{V_\DT}{V_\Di}\right)^{1/3},  \;
\av{\varrho} = \frac{M_\CD}{V_\Di\, a_\CD^{3}}
= \frac{M_\CD}{V_\DT}(1+z)^3 \, .
 \label{eq:norm}
\end{equation}
Spatially, the approximation $a(t)=\at (t)$ implies that the Friedmannian spatial volume
differs from the volume of the averaged inhomogeneous domain by a factor $\sqrt{\det (h_{ij})}$ that contains metrical
deviations from smooth (constant-curvature) space sections and is overall volume-weighted. For large-scale averages well beyond objects like neutron stars or black holes and especially for volumes as large as the DESI volume this difference is less important than the derivatives of the metric, notably the curvature that is quantitatively highly important, i.e. metrical perturbations of order $10^{-6}$ are expected to result in curvature deviations of order $1$ \cite{curvatureestimates}.

For the representation in terms of redshift the situation is different. R\"as\"anen showed that for a statistically homogeneous and isotropic dust universe with
vanishing vorticity and a well-defined homogeneity scale, the null
geodesic bundle samples the structures in proportion to their volume fraction as a function of an
\textit{average} redshift $\bar z$ (simply denoted by $z$ in \eqref{eq:rasanen} and throughout).

R\"as\"anen's theorem is a statement about averages on the light cone, so
it carries potential caveats worth keeping in mind. It is only here where an improvement of the approximation 
could result in an improvement of our closure strategy. We mention three main points.
\begin{enumerate}\itemsep2pt

\item[a)] \textit{Fair sampling.} The proof requires that the expansion rate, averaged along the light bundle, 
equals the volume average. This is an ergodic statement that requires a homogeneity scale and the expectation
that fluctuations between the two averages are small on that scale.
That light spends disproportionately much time in underdense regions is not by itself a bias, since
voids dominate the volume and this is already contained in $H_\CD$, being also compensated by a correspondingly opposite effect in overdense regions. A bias would
require a selection of rays, for instance because sources sit in overdense
regions. Such an effect concerns the sampling, not the volume-averaged dynamics. An eventual net effect 
is equally ignored in Friedmannian modeling.

\item[b)] \textit{Averages vs. individual rays and distances.} Equation~\eqref{eq:rasanen} holds for an average redshift $\bar z$; the scatter of individual lines of sight is a separate (and observationally accessible)
quantity. Only the mean is being matched to data. 
The companion result for the angular diameter distance additionally requires the
null shear/Weyl focusing contribution to be negligible. It enters the Sachs equation
quadratically, so it does not cancel between structures of different orientation. What is 
needed is therefore a bound on its magnitude. The shear term enters
through $D_A/{\bar D}_A=\sqrt{\det A}\simeq1-\kappa-\frac12|\gamma|^2$, with the lensing convergence 
$\kappa$ and the lensing shear $\gamma$, so the mean
distance is controlled by the convergence variance $\langle\kappa^2\rangle$, which for a
statistically isotropic field equals the total shear variance measured by weak lensing.
With the second-order lensing estimate of \cite{genevaCMB},
$\frac32\langle\kappa^2\rangle\simeq0.014\,(\Omega_m h^2/0.14)^3(\chi/14\,\mathrm{Gpc})^3$,
this gives $\langle\kappa^2\rangle\simeq3\times10^{-5}$--$7\times10^{-4}$ over the DESI
range, so the induced bias on $\bar D_A$ stays at the
$0.1\,\%$ level, which is below the focusing mismatch quantified in Appendix~\ref{distances}.

\item[c)] \textit{Transverse distances.} Closure is worked out with the inputs of the same expansion history and the same angular diameter distances as functions of the redshift as adopted by the DESI team,
which uses the Baryonic Acoustic Oscillations (BAO) constraints on the transverse comoving distance $D_M$ with respect to the sound horizon $r_d$ at the drag epoch, $D_M (z) / r_d$, with $D_M (z) = c \int_0^z dz' / H(z')$; the Hubble distance is denoted by $D_H (z)= c/H(z)$. The $D_M$ calibration has to be put under scrutiny. We will comment on this in Appendix~\ref{distances} 
and show that the reconstruction of $\CQ_\CD$ and $\CW_\CD$ from the radial BAO data alone involves no light-propagation model beyond $1+z=a_{\DT}/a_\CD$ (modulo the BAO compression itself, and that the backreaction variables computed with no transverse data match our benchmark.\\
\end{enumerate}

Strictly, the relevant object is a light-cone
average \cite{gasperini2010, gasperini2011}, \cite{lightcone}, and \eqref{eq:rasanen} 
amounts to interchanging it with the spatial average on constant-$t$ hypersurfaces. 
R\"as\"anen's proof requires the distribution of structures to evolve
slowly compared with the light-crossing time of the homogeneity scale, and the ray to be
averaged over distances longer than that scale but much shorter than the Hubble scale.
Over the much longer light-crossing time of the whole domain the geometry does evolve,
and the residual non-commutativity is the actual content of the approximation.

A thorough discussion of the role of the averaged expansion rate and its relation to redshift and distances within an exact Swiss-Cheese construction called ``Tardis spacetime'' has been provided in \cite{tardis}, and a perturbative analysis 
in \cite{rasanen2012}.
Complementary is a study of light propagation in exact inhomogeneous toy-models that do not use a FLRW background for the test of R\"as\"anen's approximation \cite{koksbangtest}, 
see also a test on the comparison of line-of-sight averages with volume averages \cite{koksbang1Dtest}, and the analysis of Appendix~\ref{distances}. 

The adopted approximation remains an idealization and challenges further refinement.\footnote{An alternative phenomenological closure within the \textit{Timescape model} without using this approximation has been investigated by Wiltshire \cite{wiltshire:observations}.}

\subsection{Closing the system (\ref{effectivefriedmann})}

\begin{figure*}\centering
\includegraphics[width=0.8\textwidth]{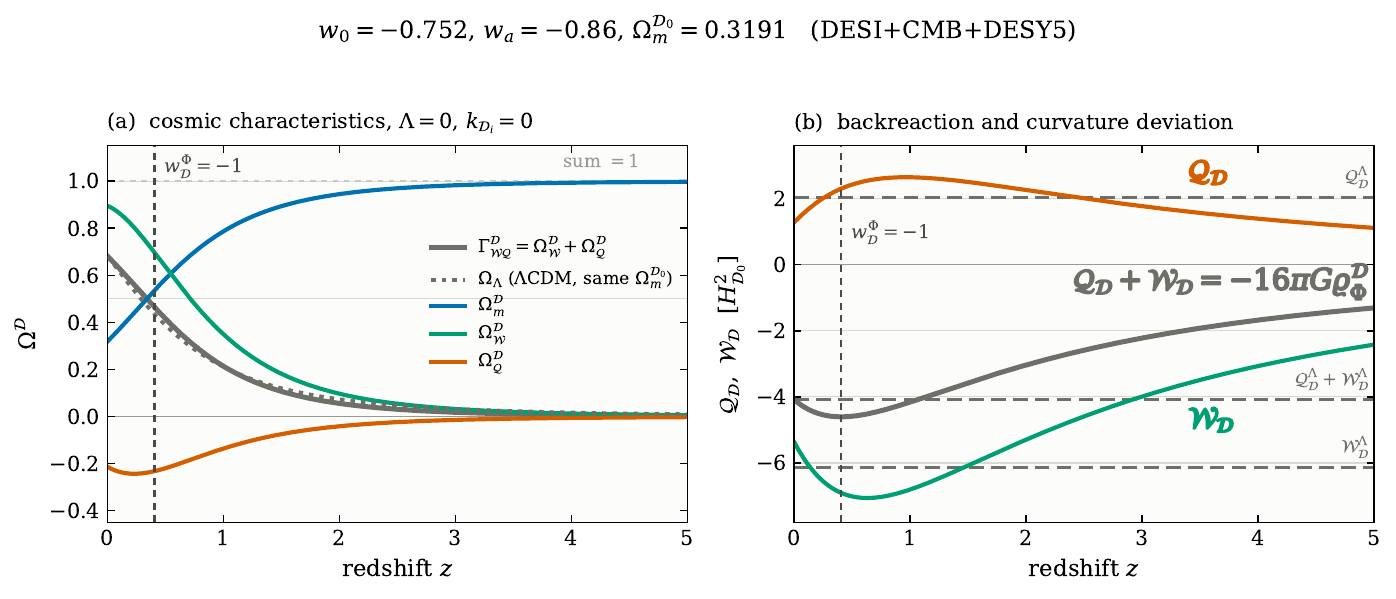}
\captionsetup{width=0.9\linewidth, format=hang}
\caption{(a) the cosmic characteristics $\Omega_m^\CD,\Omega_\CW^\CD,
\Omega_\CQ^\CD$ for $\Lambda=0$, $k_\Di=0$ for the CPL parameters on top; their sum equals $1$.
(b) $\CQ_\CD$ and $\CW_\CD$ in units of $H_\DT^2$. The grey full line in both panels depicts the sum representing 
the dark energy content mimicked by the geometric variables, compared with $\Omega_\Lambda$  in the standard model (dotted grey). The dashed vertical line marks
$w^\Phi_\CD=-1$ (or the phantom divide at $z_\times=0.405$), where $3\CQ_\CD+\CW_\CD=0$.}
\label{fig:omegas}
\end{figure*}

We set $\Lambda=0$ and $k_\Di=0$ throughout (with $k_\Di=0$, $\CW_\CD=\av{\CR}$). The averaged
energy constraint \eqref{averagehamilton}, $3H_\CD^2=8\pi G(\av{\varrho}+\rhoP)$, 
eliminates one of the two geometrical terms in favour of $H_\CD$ and the domain's restmass:
\begin{equation}
 \rhoP=-\frac{\CQ_\CD+\CW_\CD}{16\pi G}=\frac{3H_\CD^2}{8\pi G}-\frac{M_\CD}{V_\DT \at^3}\; .
 \label{eq:elim}
\end{equation}
Now, we impose the \textit{closure condition}:
\begin{equation}
w_\CD^\Phi (\at)\;\stackrel{!}{=}\;w_{\rm DESI}(\at)\;=\;w_0+w_a(1-\at)\; .
 \label{eq:closure}
\end{equation}
Since $\rhoP$ obeys $\dot{\rhoP}+3H_\CD(1+w^\Phi_\CD)\rhoP=0$, which is the integrability condition \eqref{integrability},
the closure integrates:
\begin{equation}
 \rhoP(\at)=\rhoP\big|_{0}\;f(\at)\; ,
 \quad
 f(\at)=\at^{-3(1+w_0+w_a)}\,e^{-3w_a(1-\at)} \;.
 \label{eq:rhoPhi}
\end{equation}
Evaluating \eqref{eq:elim} at the present epoch fixes the amplitude:
\begin{subequations}
\begin{align}
 \rhoP\big|_{0}&=\frac{3H_\DT^2}{8\pi G}-\frac{M_\CD}{V_\DT}
 =\frac{3H_\DT^2}{8\pi G}\left(1-\Omega_m^{\DT}\right) \; ,\\
 \Omega_m^{\DT}&:=\frac{8\pi G\,M_\CD}{3H_\DT^2\,V_\DT}
 =\frac{8\pi G\,M_\CD}{3H_\DT^2\,V_\Di\,a_\DT^{3}} \; .
 \label{eq:amp}
\end{align}
\end{subequations}
We are now in the position to evaluate $\CQ_\CD$ and $\CW_\CD$ explicitly.
Equations $\{ \eqref{eq:elim}, \eqref{eq:closure}\}$ form a linear $2\times2$ system for
$\{\CQ_\CD, \CW_\CD\}$:
\begin{equation}
 \CQ_\CD+\CW_\CD=-16\pi G\,\rhoP ,
 \quad
 -3\CQ_\CD+\CW_\CD=48\pi G\,w^\Phi_\CD\,\rhoP \; ,
\end{equation}
whose solution is
\begin{equation}
 \CQ_\CD=-4\pi G\left(1+3w^\Phi_\CD\right)\rhoP \; ,
 \quad
 \CW_\CD=12\pi G\left(w^\Phi_\CD-1\right)\rhoP \; .
 \label{eq:QWgeneral}
\end{equation}
Inserting \eqref{eq:rhoPhi}--\eqref{eq:amp} gives the result in closed form:
\begin{subequations}
\begin{align}
 \CQ_\CD (\at)&=-\frac{3}{2}H_\DT^2\left(1-\Omega_m^{\DT}\right)
   \Big[1+3w_0+3w_a(1-\at)\Big]\,\times \nonumber\\
   &\quad \ \,\at^{-3(1+w_0+w_a)}e^{-3w_a(1-\at)} \; ,
   \label{eq:Qfinal}\\[4pt]
 \CW_\CD(\at)&=+\frac{9}{2}H_\DT^2\left(1-\Omega_m^{\DT}\right)
   \Big[w_0-1+w_a(1-\at)\Big]\,\times\nonumber\\
  &\quad \ \, \at^{-3(1+w_0+w_a)}e^{-3w_a(1-\at)} \; ,
   \label{eq:Wfinal}
\end{align}
\end{subequations}
with $\at=a_\CD / a_\DT$. Equations~\eqref{eq:Qfinal} and \eqref{eq:Wfinal} are the CPL 
versions of the general local equations \eqref{Edependence} given in Appendix~\ref{nonparametric} 
and exploited there for a nonparametric closure.
In terms of redshift, using \eqref{eq:rasanen}, we have for CPL:
\begin{subequations}
\begin{align}
 f(z)&=(1+z)^{3(1+w_0+w_a)}\exp\!\left[-3w_a\frac{z}{1+z}\right],\\
 w^\Phi_\CD(z)&=w_0+w_a\frac{z}{1+z} \; ,
\end{align}
\end{subequations}
and explicitly in terms of the mass, $\tfrac32 H_\DT^2(1-\Omega_m^{\DT})
=\tfrac32H_\DT^2-4\pi G M_\CD/V_\DT$, we obtain:
\begin{subequations}
\label{aD0dropped}
\begin{align}
\CQ_\CD &=-\left(\frac{3H_\DT^2}{2}-\frac{4\pi G M_\CD}{V_\DT}\right)\big[1+3w^\Phi_\CD\big]f \; ,\\
\CW_\CD &= 3\left(\frac{3H_\DT^2}{2}-\frac{4\pi G M_\CD}{V_\DT}\right)\big[w^\Phi_\CD-1\big]f\; .
\end{align}
\end{subequations}
Note that $a_\DT$, and with it the initial normalization $a_\CD (t_{\rm i})=1$, has dropped out
of every observable: it survives only inside $V_\DT=V_\Di a_\DT^3$, i.e.\ inside
$\Omega_m^{\DT}$. The ratio of the two geometrical terms is a 
function of $w^\Phi_\CD$, independent of the amplitude and hence of $f$:
\begin{equation}
 \frac{\CQ_\CD}{\CW_\CD} = -\frac{1+3 w^\Phi_\CD}{3\,(w^\Phi_\CD-1)} \ ;\ \ w^\Phi_\CD = - \frac{\frac{1}{3} - \frac{\CQ_\CD}{\CW_\CD}}{1 + \frac{\CQ_\CD}{\CW_\CD}} \; .
 \label{eq:ratio}
\end{equation}
It is worth recalling here that $w^\Phi_\CD=-1/3$ gives $\CQ_\CD=0$ (pure curvature);
$w^\Phi_\CD=+1$ gives $\CW_\CD=0$ (pure, stiff backreaction), and $w^\Phi_\CD=-1$ gives
$\CQ_\CD/\CW_\CD=-1/3$, i.e.\ $3\CQ_\CD+\CW_\CD=0$, the \textit{phantom divide}, where the morphon kinetic term vanishes. 
For the baseline fit the ratio runs from $-0.239$ at $z=0$ to $-0.489$ as $z\to\infty$, the equation of state from 
$-0.752$ at $z=0$ to $-1.612$ as $z\to\infty$, \textit{cf.} Figure~\ref{fig:omegas}.

\begin{figure*}\centering
\includegraphics[width=0.8\textwidth]{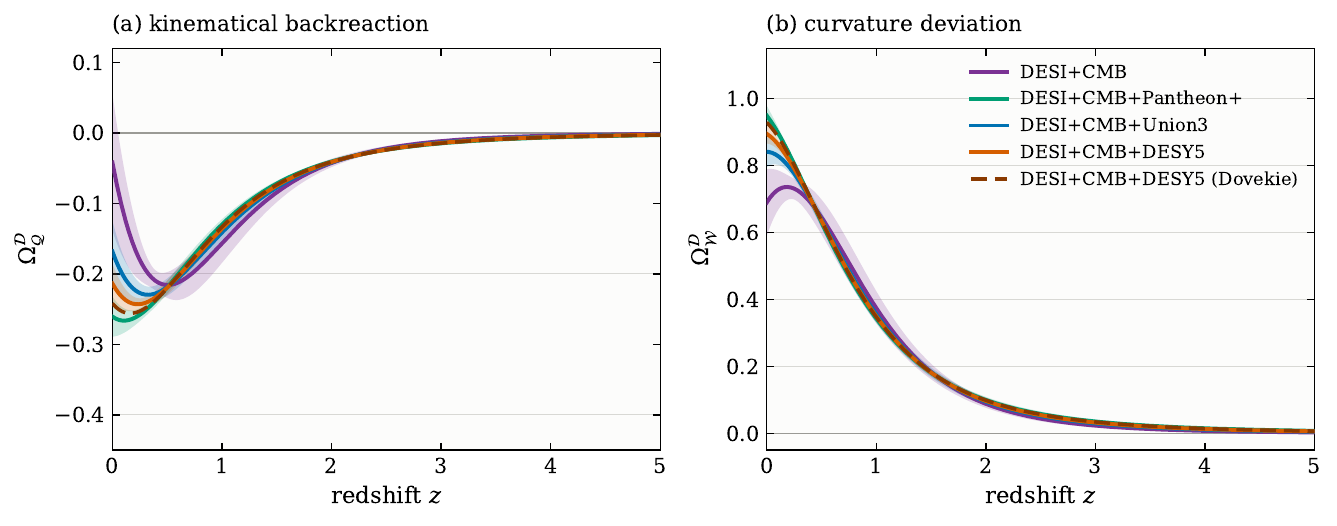}
\caption{Evolution of $\Omega_\CQ^\CD$ and $\Omega_\CW^\CD$ resulting from the closure in terms of the DESI DR2 data combinations. Bands are $68\,\%$ credible regions from the DESI DR2 uncertainties on
$(w_0,w_a,\Omega_m^{\DT})$ of each combination, propagated as in Appendix~\ref{uncertainties}; the correlation
$\rho(w_0,w_a)$ is derived from $\sigma[w(z{=}0.5)]$, and set to
$-0.91$ for Union3, to which the band is insensitive. The four combinations are
consistent at every redshift; DESI+CMB differs at $z=0$ only because it is the least
constrained. The dashed curve is the DES-SN5YR sample recalibrated with the Dovekie cross-calibration (last
row of Table~\ref{tableI}), a variant of the DESY5 combination with nearly the same uncertainties, and is therefore drawn without a band). It departs from DESY5 by $1.0\sigma$ at $z=0$ and at the pivot and by less than
$0.4\sigma$ above $z\simeq0.5$.
}
\label{fig:datasets}
\end{figure*}

\begin{table*}\centering
\begin{tabular}{@{}lrrrrrr@{}}
\toprule
Data combination & $w_0$ & $w_a$ & $\Omega^\DT_m$ & $\Omega_\CQ^\DT$ & $\Omega_\CW^\DT$ & $\CQ_\CD/H_\DT^2$ \\
\midrule
DESI+CMB                    & $-0.42$  & $-1.75$ & $0.3530$ & $-0.042$ & $0.689$ & $+0.25$\\
DESI+CMB+Pantheon+          & $-0.838$ & $-0.62$ & $0.3114$ & $-0.261$ & $0.949$ & $+1.56$\\
DESI+CMB+Union3             & $-0.667$ & $-1.09$ & $0.3275$ & $-0.168$ & $0.841$ & $+1.01$\\
DESI+CMB+DESY5              & $-0.752$ & $-0.86$ & $0.3191$ & $-0.214$ & $0.895$ & $+1.28$\\
DESI+CMB+DESY5 (Dovekie)    & $-0.803$ & $-0.72$ & $0.3130$ & $-0.242$ & $0.929$ & $+1.45$\\
\bottomrule
\end{tabular}
\caption{Present-day values implied by the closure. $\Omega_m^{\DT}$ as in the
corresponding DESI fit; $\Omega_\CQ^{\DT}+\Omega_\CW^{\DT}=1-\Omega_m^{\DT}$ by construction.
The last row uses the DES-SN5YR sample recalibrated with the Dovekie cross-calibration \cite{popovicDES}, 
which lowers the evidence for evolving dark energy from $4.2\sigma$ to $3.2\sigma$ while moving the closure's backreaction amplitude slightly upwards; the phantom divide moves to $z_\times=0.377$.}
\label{tableI}
\end{table*}

\subsection{Dimensionless form and results}

The adimensional characteristics \eqref{omega} provide the energy budget of the averaged inhomogeneous model.
With $\Lambda=0=k_\Di$ and \eqref{eq:Qfinal}--\eqref{eq:Wfinal} we obtain:
\begin{subequations}
\begin{align}
 \Omega_m^\CD(z)&=\frac{\Omega_m^{\DT}(1+z)^3}{E^2(z)} , \\[3pt]
 \Omega_\CQ^\CD(z)&=\frac{\big(1-\Omega_m^{\DT}\big)\,f(z)}{4\,E^2(z)}
   \Big[1+3w^\Phi_\CD(z)\Big] , \\[3pt]
 \Omega_\CW^\CD(z)&=\frac{3\big(1-\Omega_m^{\DT}\big)\,f(z)}{4\,E^2(z)}
   \Big[1-w^\Phi_\CD(z)\Big] ,
\end{align}
\end{subequations}
with $E^2(z):=\Omega_m^{\DT}(1+z)^3+(1-\Omega_m^{\DT})f(z)$. It can be readily verified that the sum is equal to $1$.

In terms of the deceleration functional, $q_\CD:=-\ddot a_\CD/(a_\CD H_\CD^2)=-1+\tfrac12\,\mathrm{d}\ln H_\CD^2/\mathrm{d}\ln(1+z)$,
Eqs.~\eqref{averageraychaudhuri},~\eqref{averagehamilton} provide, for $\Lambda=0=k_\Di$:
\begin{equation}
\label{eq:cosmo}
\Omega^\CD_\CQ=\frac{q_\CD}{2}-\frac{\Omega^\CD_m}{4} , \ 
\Omega^\CD_\CW=1-\frac{q_\CD}{2}-\frac{3\,\Omega^\CD_m}{4}, \ 
w^\Phi_\CD=\frac{2q_\CD-1}{3\,(1-\Omega^\CD_m)}\ .
\end{equation}
Figure~\ref{fig:omegas} shows the redshift-dependence of the dimensionless characteristics, together with 
a comparison of the bare variables. Kinematical backreaction is positive (the corresponding adimensional variable is negative) throughout all redshifts, i.e. the averages are expansion-variance dominated, while the average scalar curvature is negative (the corresponding adimensional variable is positive) throughout all redshifts.  These results are qualitatively expected: at low redshifts the volume of the Universe is void-dominated, hence a dominating negative average scalar curvature is expected, while the subdominance of average shear, hence expansion-variance dominance, is expected for a large-scale isotropic matter distribution. 

In order to see the dependence of these two characteristics on the various combinations in the DESI DR2 data, DES-SN5YR \cite{DES,DES-SN}, Pantheon+ \cite{pantheon} and Union3 \cite{union3}, we provide Figure~\ref{fig:datasets} together with Table~\ref{tableI}.

\subsection{The morphon and the phantom divide}

\begin{figure*}\centering
\includegraphics[width=0.85\textwidth]{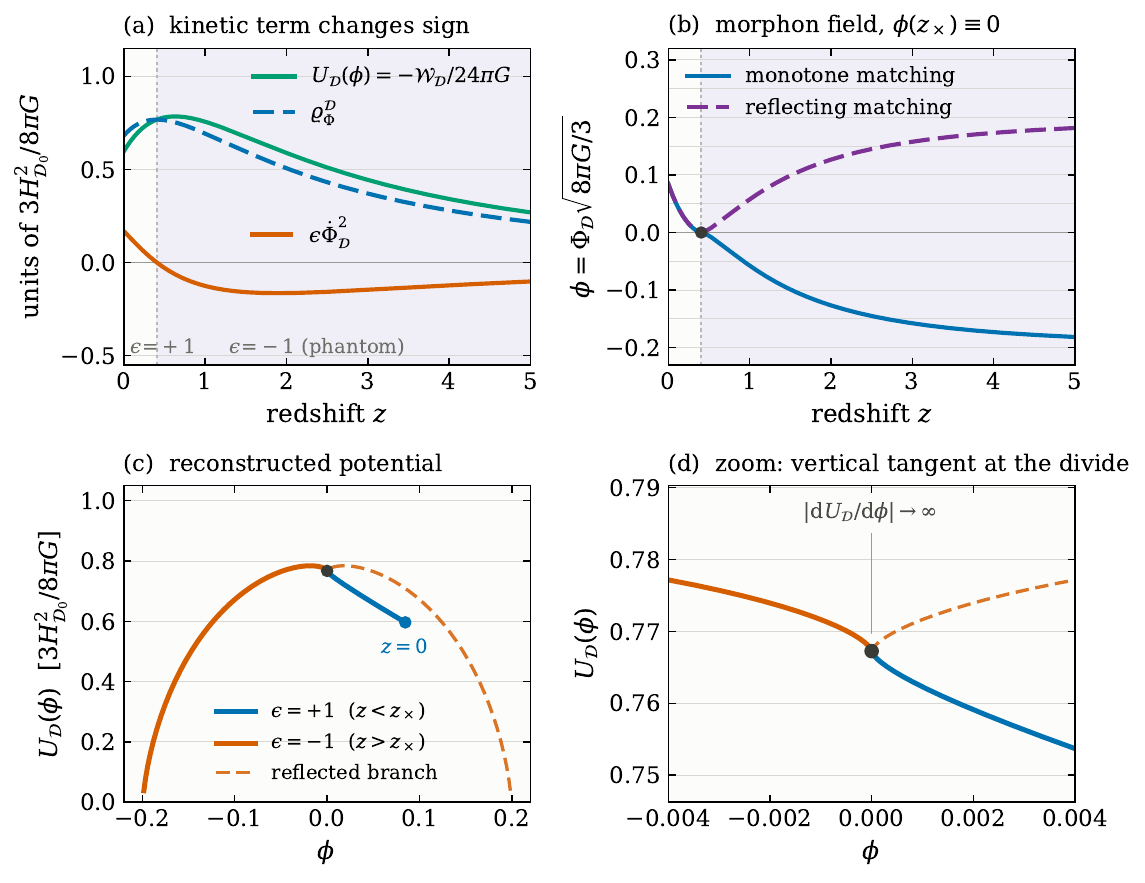}
\caption{The morphon. (a) $U_{\CD} (\phi)$, $\rhoP$ and $\epsilon{\dot\Phi}_{\CD}^2$ in units of
$3H_\DT^2/8\pi G$; the kinetic term crosses zero at $z_\times$, where $U_\CD (\phi)=\rhoP$, 
and where $\varrho^{\CD}_{\Phi}$ is maximal, since $\dot\varrho^{\CD}_{\Phi}\propto(1+w^{\Phi}_\CD)$.
(b) the rescaled field $\phi$ for the two matchings. 
(c) the reconstructed potential $U_{\CD}(\phi)$: the solid branches are cut at
$z=20$, while the standard branch ends at the present epoch, marked by $z=0$. (d) its zoom
at the divide, showing the vertical tangent.}
\label{fig:morphon}
\end{figure*}

Here, we provide the morphonic representation of the results in order to introduce the scalar field language that is usually adopted in standard model quintessence approaches. From Section~\ref{morphonic} we have:
\begin{subequations}
\begin{align}
\epsilon\,\dot\Phi_\CD^2&=\big(1+w^\Phi_\CD\big)\rhoP=-\frac{3\CQ_\CD+\CW_\CD}{24\pi G}\; ;\\
U_\CD (\Phi_\CD) &=\tfrac12\big(1-w^\Phi_\CD\big)\rhoP=-\frac{\CW_\CD}{24\pi G}\;>\;0 \;.
\end{align}
\end{subequations}
Since $\rhoP>0$, the sign of the kinetic term is fixed by $w^\Phi_\CD$,
\begin{equation}
 \epsilon=\mathrm{sgn}\big(1+w^\Phi_\CD\big)
 =\begin{cases}
 -1 & \tilde a<\tilde a_\times \quad(z>z_\times):\ \text{phantom},\\[2pt]
 +1 & \tilde a>\tilde a_\times \quad(z<z_\times):\ \text{standard},
 \end{cases}
\end{equation}
with $\tilde a_\times =1+\frac{1+w_0}{w_a}$.
Thus, the DESI closure requires \textit{two} morphon branches, an early phantom phase and
a present standard phase (the ``quintom-B'' ordering \cite{quintom}), matched at $z_\times=0.405$.
Note that the morphon's potential energy is the emerging negative average curvature alone, while its kinetic energy is the
combination $3\CQ_\CD+\CW_\CD$.

\subsubsection{Reconstruction}

Changing the time-derivative variable $t$ to $\at$, ${\dot\Phi}_\CD=\tilde a H_\CD\,\mathrm{d}\Phi_\CD/\mathrm{d}\tilde a$ and defining the rescaled field $\phi:=\Phi_\CD \sqrt{8\pi G/3}$ ($\Phi_\CD=\sqrt3\,M_{\rm Pl}\,\phi$), used from here on for the potential and the field trajectory with the shorthand $U_{\CD}(\phi):=U_\CD(\Phi_\CD(\phi))$, we obtain:
\begin{subequations}
\begin{align}
 &\frac{\mathrm{d}\phi}{\mathrm{d}\tilde a}
 =\pm\,\frac{\sqrt{\big(1-\Omega_m^{\DT}\big)\,\epsilon (1+w^\Phi_\CD(\tilde a))\,f(\tilde a)}}
 {\tilde a\,E(\tilde a)} \; ,\\
 &\frac{8\pi G\,U_\CD(\phi)}{3H_\DT^2}=\tfrac12\big(1-w^\Phi_\CD\big)\big(1-\Omega_m^{\DT}\big)f(\tilde a) \;.
 \label{eq:phirec}
\end{align}
\end{subequations}
The integral converges at both ends; for the baseline the total field excursion over
$0<z<\infty$ is $\Delta\phi\simeq0.28$, i.e.\ $\Delta\Phi_\CD\simeq0.49\,M_{\rm Pl}$. 

\subsubsection{Matching conditions at the divide}

Continuity of $\phi$, ${\dot\phi}$ and $U_\CD(\phi)$ is required. 
$U_\CD(\phi)=\tfrac12(1-w^\Phi_\CD)\rhoP$ is manifestly continuous, and also ${\dot\phi} \to0$ on both sides, so the kinetic matching is trivially satisfied. At $\tilde a_\times$ one has $U_\CD =\rhoP=-\pP$. What is not trivial is the geometry of $U_\CD$ as a function of $\phi$. 
Near the divide $1+w^\Phi_\CD\simeq-w_a(\tilde a-\tilde a_\times)$, so that
\begin{subequations}
\begin{align}
& \phi-\phi_\times\simeq\pm\frac{2}{3}\,
 \frac{\sqrt{(1-\Omega_m^{\DT})| w_a | f_\times}}{\tilde a_\times E_\times}\,
 \big|\tilde a-\tilde a_\times\big|^{3/2} \;,\\
 & \left|\frac{\mathrm{d}U_\CD (\phi)}{\mathrm{d}\phi}\right|
 \;\propto\;\big|\tilde a-\tilde a_\times\big|^{-1/2}\;\longrightarrow\;\infty \; ,
\end{align}
\end{subequations}
because $\mathrm{d}U_\CD /\mathrm{d}\tilde a\neq0$ there, \textit{cf.} Figure~\ref{fig:morphon}. 
The reconstructed potential has a vertical tangent at the divide
(Fig.~\ref{fig:morphon}d). This is the familiar obstruction to phantom crossing $w=-1$ with one canonical
scalar, here appearing as a harmless kink in a bookkeeping variable.

The residual freedom is the sign of ${\dot\phi}$ on each branch, giving two inequivalent matchings:
{\small 
\begin{itemize}\itemsep2pt
\item[(i)] \textit{monotone}: ${\dot\phi} >0$ throughout. $\phi$ increases through
 $\phi_\times$ with a momentary halt; $U_\CD (\phi)$ is single-valued with a vertical
 inflection.
\item[(ii)] \textit{reflecting}: ${\dot\phi}$ changes sign at the divide. $\phi$ has a
 turning point; $U_\CD (\phi)$ becomes double-valued, two sheets meeting at a vertical
 tangent.
\end{itemize}
}
Both reproduce identical $\CQ_\CD,\CW_\CD,H_\CD$, the choice being pure gauge of the morphon
correspondence.

\subsection{The averaging domain and the homogeneity scale}

An open question is the dependence on spatial scale, represented by the domain index $\CD$ in all formulae, 
while a Friedmannian-type fit applies to all scales. We are going to look at the domain-dependence
of our analysis using a multiscale partitioning approach that was developed for the general case in \cite{buchertcarfora2008} and \cite{wiegand2010}, see also \cite{wiltshire:twoscale,wiltshire:observations,wiltshire:dust}
and the recent work \cite{partitioning}.

\subsubsection{Exact partition formulae}

Let $\CD=\bigsqcup_\ell\CD_\ell$ be a disjoint partition into subdomains with volume
fractions $\lambda_\ell=V_{\CD_\ell}/V_\CD$, $\sum_\ell\lambda_\ell=1$. Additive volume averages of
scalars, $\av{\Psi}=\sum_\ell\lambda_\ell\langle\Psi\rangle_{\CD_\ell}$, 
are $\av{\varrho}$, $\av{\CR}$, $\av{\sigma^2}$, and
$H_\CD=\sum_\ell\lambda_\ell H_{\CD_\ell}$. However, inserting this into \eqref{eq:Q-GR} yields:
\begin{equation}
 \CQ_\CD = \;\sum_\ell\lambda_\ell\,\CQ_{\CD_\ell}
 \;+\;3\sum_{\ell\neq m}\lambda_\ell\lambda_m
 \big(H_{\CD_\ell}-H_{\CD_m}\big)^2\; .
 \label{eq:partQ}
\end{equation}
Thus, $\CW_\CD$ is additive (for $k_{\CD_i} = 0$), while $\CQ_\CD$ is strictly superadditive: coarse-graining
never destroys backreaction, it creates it through the volume-weighted variance of
the regional Hubble rates. This is the precise sense in which $\CQ_\CD$ is
domain-dependent and the reason a closure condition must specify the domain.

\subsubsection{$\lambda_H$ as the saturation scale of $\CQ_\CD$}

Equation \eqref{eq:partQ} gives an operational, backreaction-intrinsic definition of the
homogeneity scale. For $L_\CD=V_\CD^{1/3}$ below $\lambda_H$, enlarging $\CD$ adds new
regional Hubble contrasts and $\CQ_\CD (L_\CD)$ grows; once $L_\CD\gtrsim\lambda_H$ the
variance term saturates and $\CQ_\CD$ becomes domain-independent up to sampling
fluctuations. Treating the domain as $N=(L_\CD/\lambda_H)^3$ quasi-independent cells, we 
write \eqref{eq:partQ} in the form:
\begin{align}
\CQ_\CD &= \frac{1}{N}\sum_\ell \CQ_{\CD_\ell} + 6\,S^2_N(H)\ , \ \textrm{with}\nonumber\\
S^2_N(H):&=\frac{1}{N}\sum_\ell H_{\CD_\ell}^2-\Big(\frac{1}{N}\sum_\ell H_{\CD_\ell}\Big)^2 \, ,
\end{align}
where $\ 3\sum_{\ell\neq m}\lambda_\ell\lambda_m (H_{\CD_\ell}-H_{\CD_m})^2 = 6\,S^2_N(H)$.\\

\begin{figure*}\centering
\includegraphics[width=0.9\textwidth]{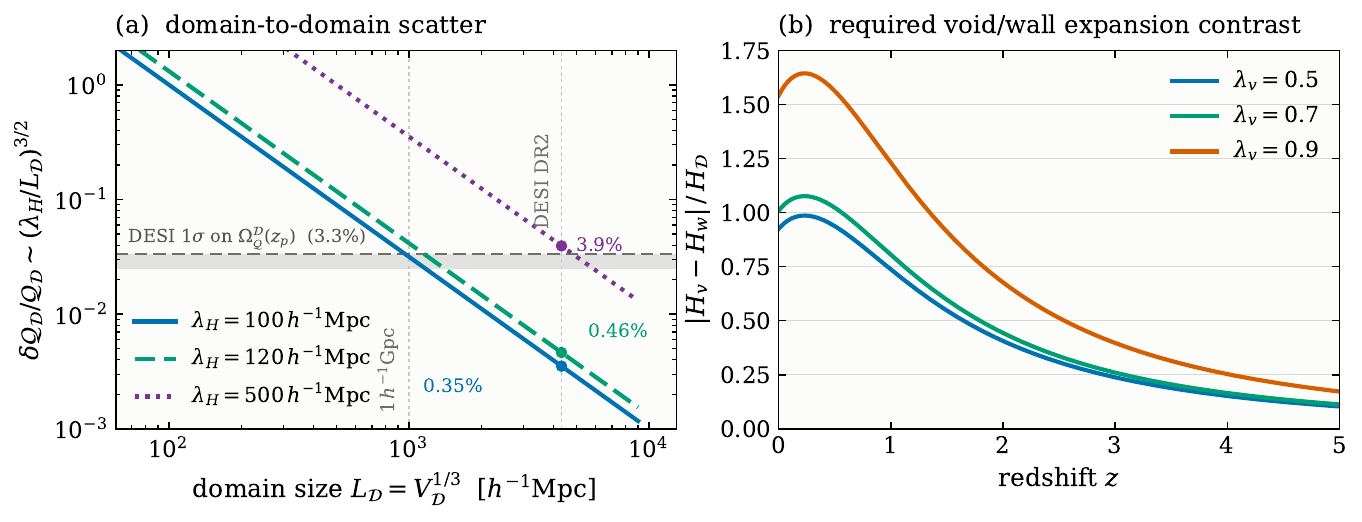}
\captionsetup{width=0.73\linewidth, format=hang}
\caption{(a) the central-limit estimate \eqref{eq:cosmicvar} of the domain-to-domain scatter of $\CQ_\CD$.
(b) the bound \eqref{eq:bound}, conditional on $\lambda_v \CQ_v+\lambda_w \CQ_w \le 0$, in units of $H_\CD(z)$.}
\label{fig:domain}
\end{figure*}

\begin{table*}\centering
\begin{tabular}{lccc|ccc}
\toprule
 & \multicolumn{3}{c|}{$|H_v-H_w|/H_\CD$} & \multicolumn{3}{c}{$H_v/H_\CD\ ;\ H_w/H_\CD$} \\
$z$ & $\lambda_v=0.5$ & $0.7$ & $0.9$ & $\lambda_v=0.5$ & $0.7$ & $0.9$\\
\midrule
$0$ & $0.93$ & $1.01$ & $1.54$ & $1.46\,;\,0.54$ & $1.30\,;\,0.29$ & $1.15\,;\,-0.39$\\
$0.5$ & $0.94$ & $1.02$ & $1.56$ & $1.47\,;\,0.53$ & $1.31\,;\,0.28$ & $1.16\,;\,-0.41$\\
$1$ & $0.74$ & $0.80$ & $1.23$ & $1.37\,;\,0.63$ & $1.24\,;\,0.44$ & $1.12\,;\,-0.11$\\
$2$ & $0.41$ & $0.44$ & $0.68$ & $1.20\,;\,0.80$ & $1.13\,;\,0.69$ & $1.07\,;\,+0.39$\\
\bottomrule
\end{tabular}
\captionsetup{width=0.93\linewidth, format=hang}
\caption{Minimal void/wall expansion contrast implied by the DESI closure, from
\eqref{eq:bound}. At $\lambda_v=0.9$ the walls must already be collapsing on average out to $z\simeq1$. 
(Beware that the volume fractions are time-dependent, so the $z\ge1$ rows are only illustrative.)}
\label{tableII}
\end{table*}

The homogeneity scale is conventionally defined so that the deviation from homogeneity
smoothed on that scale is a few percent (the fractal correlation dimension
reaching $D_2=3$, or the scaled counts-in-spheres reaching unity, to within about one
percent). With an rms density contrast $\delta\simeq0.01$--$0.03$ on $\lambda_H$, and the
linear relation\footnote{For this estimate, we adopt the simple linear theory relation on a FLRW background. Strictly, this contradicts our statement that the background is not FLRW: the estimate is an approximation dropping fluctuation sources from the general relation of perturbations on the average distribution investigated in \cite{roy2012}.} between the cell-averaged expansion rate and the enclosed overdensity,
$\theta^{\rm pec}=-aH \xi\delta$, i.e.\ $\delta H/H\simeq-\frac13 \xi \delta$ with the growth rate 
$\xi=(\Omega_m^{\DT})^{0.55}=0.534$, one has $\delta H/H\simeq0.2$--$0.5\,\%$.  Here, 
${\rm Var}(H_{\CD_\ell})$ is the variance of the cell-averaged rates across cells, so that
\begin{equation}
\frac{6\,{\rm Var}(H_{\CD_\ell})}{H^2_{\DT}}\simeq 2\times10^{-5}\!-\!2\times10^{-4}
\ \text{against}\  \frac{\CQ_\CD}{H^2_{\DT}}\simeq 1.28 \; .
\end{equation}
We conclude that, on the homogeneity scale, the dominant term is the volume-weighted mean of the regional backreaction terms, not the variance, and we can estimate the standard deviation of the total backreaction, assuming now equal cells, $\lambda_\ell \approx 1/N$, as
\begin{equation}
 \frac{\delta\CQ_\CD}{\CQ_\CD}\;\sim\;N^{-1/2}=\left(\frac{\lambda_H}{L_\CD}\right)^{3/2},
 \label{eq:cosmicvar}
\end{equation}
with the relative cell-to-cell dispersion of $\CQ_{\CD_\ell}$ set to unity.

To give some numbers: 
with the comoving distance to the survey edge for CPL,
$\chi(z{=}2.33)=3839\,h^{-1}$Mpc, and the DR2 sky coverage
($\sim$14\,000\,deg$^2$, a fraction $f_{\rm sky}=0.339$ of the sphere), the geometric
comoving volume (approximating the light-cone volume) is $V_\CD=(4\pi /3)\chi^3 f_{\rm sky}=80\,(h^{-1}{\rm Gpc})^3$,
which yields $L_\CD=V_\CD^{1/3}=4.32\,h^{-1}$Gpc. 
Choosing for the homogeneity scale $\lambda_H\simeq100, 120\,h^{-1}$Mpc, this gives 
for the domain-to-domain scatter $0.35\%$, $0.46\%$, \textit{cf.} Fig.~\ref{fig:domain}a. The investigated closure should therefore be read as holding on a domain of at least $\sim1\,h^{-1}$Gpc, where the domain-to-domain scatter
in $\CQ_\CD$ is below the DESI parameter uncertainties investigated in Appendix~\ref{uncertainties}. Below that,
$(w_0,w_a)$ cannot be treated as domain-independent constants.

As a disclaimer we mention that the two values of the
homogeneity scale in Fig.~\ref{fig:domain}a are chosen to match commonly quoted values. This scale is, however, very  conservative. It arises by looking at two-point statistics, while structure is measured by all orders of the correlations. Minkowski functionals cover this and lead to homogeneity scales that are substantially larger \cite{minkowski}. For example, taking the DESI volume and a homogeneity scale of $\sim0.5\,h^{-1}$Gpc results in a scatter of $3.9\%$.
The $1\,h^{-1}$Gpc scale still keeps the scatter at or below the DESI
$1\sigma$ for $\lambda_H\lesssim100\,h^{-1}$Mpc. 
(Note that the scatter enters the error budget in quadrature.)

With the aim to look at the mutual expansion difference between under- and overdense unions of domains, we specialize \eqref{eq:partQ} to two types named `voids' and `walls', $\ell\in\{v,w\}$. We have
$\lambda_v+\lambda_w=1$ and 
\begin{equation}
 \CQ_\CD=\lambda_v \CQ_v+\lambda_w \CQ_w
 +6\lambda_v\lambda_w\big(H_v-H_w\big)^2 \; .
 \label{eq:twophase}
\end{equation}
If each phase is internally shear-dominated, $\lambda_v \CQ_v+\lambda_w \CQ_w \le 0$,
Equation~\eqref{eq:twophase} turns the closure into a \textit{lower bound} on the expansion
contrast, \textit{cf.} Figure~\ref{fig:domain}b:
\begin{equation}
 \big| H_v-H_w\big| \;\ge\; \sqrt{\frac{\CQ_\CD}{6\lambda_v\lambda_w}} \; ,
 \label{eq:bound}
\end{equation}
which \textit{via} $H_\CD=\lambda_v H_v + \lambda_w H_w$ fixes both phase rates. Table~\ref{tableII} evaluates it.

This is a concrete falsifiable consequence of the closure: a $\sim30\%$
enhancement of the void expansion rate over the global domain average is consistent with what
relativistic simulations and Lema\^itre--Tolman--Bondi/Szekeres models report, e.g. \cite{GRsim}, \cite{szekeresvoids}. 
If we set the required suppression in the walls, $H_w\simeq0.3\,H_\CD$, $\lambda_v=0.7$ may be considered as a conservative falsification benchmark for the present closure modeling. This sits at the top of the reported range in \cite{GRsim} for a spatial resolution of $4\,h^{-1}$Mpc, while the reported void-rate bottoms out towards the global average if void domains intersect with walls. Note that the wall side is not currently testable: 
the wall phase is below the resolution scale and has no well-defined domain expansion rate, so the comparison is one-sided and rests on the void rate together with the partition identity $H_\CD=\lambda_v H_v+\lambda_w H_w$.
Compared with the fluid-limit scale, $\lambda_v=0.7$ is still conservative and makes their values lower limits, \textit{cf.} the summary given in Sect.~\ref{IV} A.2.

\section{Physical remarks}
\label{IV}

We here list the important results with an eye on physical justifications that would allow to interpret the outcome as a benchmark for explicit structure formation models.
We will summarize issues with current structure formation models in the next subsection.

\subsection{List of important results}

\begin{enumerate}\itemsep1pt

\item \textbf{DESI, read through the closure, requires a nonscaling solution.}
By \eqref{eq:closure} and \eqref{eq:ratio} the effective equation of state  runs from $-0.752$ at $z=0$ to $-1.612$ as
$z\to\infty$. Constant-ratio (scaling) solutions $\CQ_\CD\propto\CW_\CD\propto a_\CD^n$ give
constant $w^\Phi_\CD$ and are therefore excluded as a global description by the DESI preference for $w_a\ne0$ at the
level of this closure. 
The DESI data are, in this language, evidence that backreaction variables do \textit{not} scale with a common power of $a_\CD$. 

\item \textbf{Backreaction switches off early and kicks in at the epoch of structure formation.}
For large redshifts we have $\Omega^\CD_{\CQ,\CW}\propto(1+z)^{3(w_0+w_a)}=(1+z)^{-4.84}$ for the
baseline DESI fit, so both terms are negligible at decoupling and approach a generic scaling regime ($n=p$). 
The closure is thus consistent with a near-FLRW Early Universe: within the framework this is the physically expected state,
so it is not a strong test, but it was not imposed from the outset. Note, however, since the data stop at $z=2.33$, the particular decay rate depends on the parametrization and is not a prediction. From structure formation models,  
the generic scaling solution $n=p=-1$ models the large-scale leading mode of averaged first-order 
relativistic Lagrangian perturbation solutions, 
Sect.~4.2.2 in \cite{buchert2008}, \cite{wiegand2010}, \cite{testing}, \cite{rza2}.
Hence, a physically motivated parametrization for high redshifts would impose $w_\CD^\Phi \rightarrow -2/3$ (The JBP parametrization ($w_\CD^\Phi \rightarrow w_0, \, n = -0.74$) happens to lie close to this mode). It is also worth noting that at $z=0$, $\CQ_\CD / \CW_\CD = -0.239$ (i.e.~$w_0 = -0.752$) lies close to this leading mode, 
$\CQ_\CD / \CW_\CD =-0.2$, well away from the cosmological constant value $-1/3$.\\
Furthermore, Figure~\ref{fig:omegas} shows that the onset of a substantial
backreaction occurs around $z \sim 2-3$. This outcome derives from DESI data without model assumptions and
matches the physical expectation for backreaction to kick in at the epoch of structure formation, commonly named 
\textit{coincidence problem} referring to the unexplained impact epoch of the cosmological constant.

\item \textbf{The magnitude $\CQ_\CD \simeq 1.3$--$2.6 H^2_\DT$
on $\gtrsim$ Gpc domains is the model's falsifiable content.} 
We find $\CQ_\CD>0$ over the range where the data constrain it ($z\lesssim1.8$ non-parametrically, 
and for all $z$ under CPL).
Since $w_0<-1/3$ and $w_a<0$ for every DESI combination, $1+3w^\Phi_\CD(z)<0$ for all
$z\ge0$, hence $\CQ_\CD>0$ throughout for CPL, peaking at $\CQ_\CD\simeq2.6\,H_\DT^2$ near $z\simeq1$.
By definition of $\CQ_\CD$ this demands $\tfrac13\,\mathrm{var}_\CD(\theta)>\av{\sigma^2}$ , i.e. the variance of the 
expansion rate must dominate the average shear rate on $\CD$, at the level of the
Hubble scale. This is precisely the regime that Newtonian and perturbative treatments
(where $\CQ_\CD$ is a total divergence \cite{buchertehlers}) cannot access. In terms of our multiscale analysis, 
a void/wall partition
with the conservative value $\lambda_v=0.7$ must carry $H_v\gtrsim1.30\,H_\CD$ and $H_w\lesssim0.29\,H_\CD$ today. A partition with a higher void fraction of $\lambda_v=0.9$, say at the extreme spatial resolution limit of a fluid approximation scale, points to a lower void expansion rate, $H_v\simeq1.15\,H_\CD$, and already collapsing unions of overdense regions $H_w\simeq-0.39\,H_\CD$ today. Thus, imposing stable clustering of ``virialized'' unions of overdense regions (locally outside the irrotational dust description), 
$H_w \approx 0$, i.e. $H_v/H_\CD=1/\lambda_v$, the void expansion rate would match the reported limits from numerical simulations of \cite{GRsim} for $\lambda_v \approx 0.8$.

\item \textbf{Negative average curvature dominates the energy budget.}
We found $\CW_\CD=\av{\CR}<0$ for all $z$, with $\Omega_\CW^{\DT}\simeq0.895$ against
$\Omega_\CQ^{\DT}\simeq-0.214$. The closure therefore says that nearly the entire
``dark energy'' budget is \textit{emerging negative curvature}, while it is
\textit{kinematical backreaction} that supplies the volume acceleration \cite{rasanen2006}, \cite{boehmrasanen}. 
This is the expected
qualitative signature of structure formation in a volume that is void-dominated with isotropic large-scale average
matter distribution; it is the quantitative version of the statement that the FLRW template 
misreads emerging curvature as dark energy.

\item \textbf{Phantom crossing is natural in this effective framework.}
The DESI fit crosses $w=-1$ at $\at_\times=1+(1+w_0)/w_a$, i.e.\ $z_\times=0.405$ for
the baseline fit. For a minimally coupled scalar field this requires $\epsilon$ to flip
its sign to a negative kinetic energy, a ghost, and this is the standard objection to a
quintessence  reading of DESI, e.g. \cite{carroll2003}, \cite{vikman}, \cite{quintom}. For backreaction there is no such obstruction: the
crossing is simply the regular geometric condition $3\CQ_\CD+\CW_\CD=0$, with no additional fundamental field, and no instability attached. The morphon is a bookkeeping device of already present physics, not an additional 
\textit{fundamental} physical degree of freedom, and its ``phantom phase'' costs nothing. 

\item \textbf{A few-percent prediction for the morphon kinetic term.}
Appendix~\ref{koksbang} reports on curvature-consistency tests \cite{curvaturetest}, \cite{tests} and bounds on the morphon kinetic term based thereon \cite{koksbang}, which may be 
considered as another falsification benchmark directly from observations. The closure strategy offers a testable 
prediction for the value of the quantity $\Omega^\CD_\CW+3\,\Omega^\CD_\CQ$ and its line-of-sight integral.

\end{enumerate}

\subsection{Discussion}

We start with a recap of the theoretical guidelines that led to the result.

\begin{itemize}
\item[a)] 
The architecture of common structure formation models and simulations force the large-scale average of intrinsic spatial curvature to match the assumed background FLRW curvature evolution.
Averaged Ricci curvature should not be confused with a Friedmannian curvature (integration-)constant. Even this latter curvature is set to $0$ for all times and on all scales in the standard model, although a negative constant curvature would capture parts of the physics described in the next point.

\item[b)]
A crucial fact is that averaged intrinsic spatial curvature does not obey a conservation law as the FLRW constant curvature or for instance the averaged density.
The result is a dominance of averaged negative curvature for epochs that are dominated by voids, while high-density objects occupy a small fraction of the total volume. A repulsive dark energy-like effect is inevitably produced on large scales. This effect naturally starts to dominate at the epoch of structure formation and inherits a redshift-dependence. Dynamically, averaged curvature is scale-dependent and evolves at a different rate compared with a constant-curvature model. (Already the leading mode of Lagrangian perturbation theory predicts a scaling of $\av{\CR} \propto a_\CD^{-1}$ compared with a Friedmannian scaling $\CR \propto a^{-2}$.)

\item[c)]
Spatially averaging the scalar parts of Einstein's equations delivers exact balance laws which, in the presently investigated case of an inhomogeneous dust cosmology, only needs \textit{one} more condition to obtain a closed system. 
Generally, we think of an explicit structure formation model as a closure condition, but here we directly used observational data to provide this closure. 
\end{itemize}

Taking the results of a closure with DESI data as a benchmark for structure formation models, we are drawn to shortcomings of the architecture of the majority of current structure formation models and general-relativistic numerical simulations. 
The problem resides in point a): these models force dynamical variables to average out on an assumed FLRW background. 
A major problem arises: assuming a background makes only sense if that background describes the true average. 
It is, however, evident from the presently used background-free formalism that the average does not lead to a FLRW cosmology. Worse, the averages are driven away from the FLRW background during the whole cosmic history, at least proven for the space of scaling solutions \cite{roy:instability}. 
This fact poses the main difficulty in modeling the present findings with current structure formation models, and conclusions drawn from them, e.g.~on the magnitude of backreaction, have to be considered preliminary, while other aspects of simulations like the relative expansion rates in voids and walls may be considered indicative.

An example for the presence of substantial intrinsic curvature can be seen in reconstruction techniques from peculiar-velocity data: employing the Cosmicflows--4++ reconstruction \cite{reconstruction} one finds substantial intrinsic curvature over the whole reconstructed volume. Removing the forcing of the large-scale behavior to the boundary condition of a cosmology with a dominating dark energy content, averaged intrinsic curvature would dominate the energy budget and is then expected to comply with our benchmark.

A possible route to improve on the architecture of current models is to abandon the $3$-torus topology and work out structure formation e.g. in a spherical space featuring a global intrinsic curvature. A hint towards a hyperbolic geometry with on average negative curvature has been outlined in the consideration of black hole formation within a spherical space \cite{sphericalspace}. While a background-free formalism is always preferable (since general relativity is a background-free theory), we can assign physical sense to structure formation on a ``background'' by investigating deviations from the average model, a formalism that has been developed in \cite{roy2012}. Our benchmark thus demands nothing less than a larger paradigmatic change of cosmological models and simulations.

\textbf{Acknowledgements:}
I wish to thank Jan Ostrowski and Jenny Wagner for insightful discussions during the work on this paper, Asta Heinesen, Sofie Koksbang, Syksy R\"as\"anen and David Wiltshire for valuable remarks on the manuscript. 
This work is part of a project that has received funding from the European Research Council (ERC) under the European Union's Horizon 2020 research and innovation program (grant agreement ERC Advanced Grant 740021--ARTHUS (2017--2023), PI: TB). 
The generation of figures, tables and numerical fits to observations has  been assisted by Anthropic's Claude; 
the author takes full responsibility for the content.

\appendix

\begin{figure*}\centering
\includegraphics[width=0.9\textwidth]{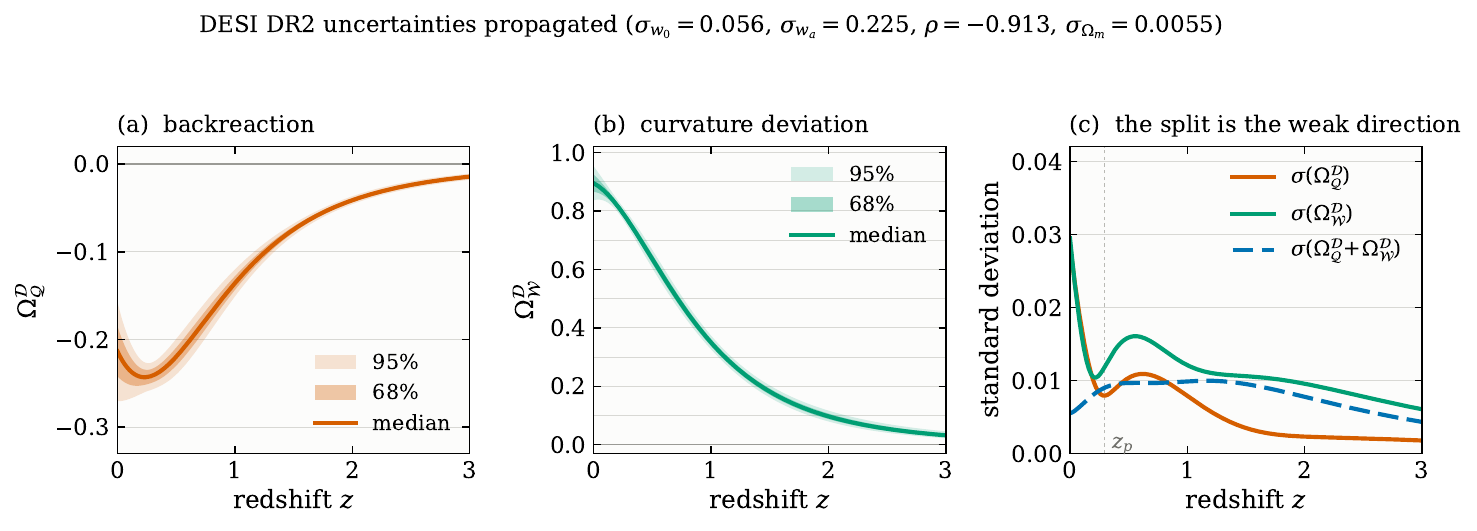}
\caption{Credible bands from the DESI DR2 covariance. Panel (c): the uncertainty on
$\Omega_\CQ^\CD+\Omega_\CW^\CD$, which equals $\sigma(\Omega_m^\CD)$ by
the sum rule; the dip near $z_p$ is the pivot.}
\label{fig:errors}
\end{figure*}

\section{Propagation of the DESI uncertainties}
\label{uncertainties}

\noindent
\textit{Analytic Jacobians:} 
Let us write $\theta\in\{w_0,w_a,\Omega_m^{\DT}\}$ and $g^\CD :=1-\Omega_m^\CD=
(1-\Omega_m^{\DT})f/E^2$ with the notations in the text, so that
\begin{equation}
\Omega_\CQ^\CD=\tfrac14 g^\CD\,(1+3w^\Phi_\CD) \ , \ 
\Omega_\CW^\CD=\tfrac34 g^\CD\,(1-w^\Phi_\CD)\; . 
\end{equation}
From \eqref{eq:f},
\begin{align}
& \frac{\partial\ln f}{\partial w_0}=3\ln(1+z)\; , \ 
 \frac{\partial\ln f}{\partial w_a}=3\ln(1+z)-\frac{3z}{1+z}\; ,\nonumber\\
& \frac{\partial w^\Phi_\CD}{\partial w_0}=1 \; , \  
 \frac{\partial w^\Phi_\CD}{\partial w_a}=\frac{z}{1+z}\; ,
\end{align}
and, after a short calculation,
\begin{align}
\frac{\partial g^\CD}{\partial\theta}&=g^\CD\,\Omega_m^\CD\,\frac{\partial\ln f}{\partial\theta}
 \quad(\theta=w_0,w_a) \ , \nonumber\\
 \frac{\partial g^\CD}{\partial\Omega_m^{\DT}}
 &=-\frac{\Omega_m^\CD\big(1-\Omega_m^\CD\big)}
 {\Omega_m^{\DT}\big(1-\Omega_m^{\DT}\big)} \; .
\end{align}
Hence,
\begin{subequations}
\begin{align}
 \frac{\partial\Omega_m^\CD}{\partial\theta}&=-\frac{\partial g^\CD}{\partial\theta}\; ,\\
 \frac{\partial\Omega^\CD_\CQ}{\partial\theta}
 &=\frac{1+3w^\Phi_\CD}{4}\frac{\partial g^\CD}{\partial\theta}
 +\frac{3g^\CD}{4}\frac{\partial w^\Phi_\CD}{\partial\theta}\; ,\\
 \frac{\partial\Omega^\CD_\CW}{\partial\theta}
 &=\frac{3(1-w^\Phi_\CD)}{4}\frac{\partial g^\CD}{\partial\theta}
 -\frac{3g^\CD}{4}\frac{\partial w^\Phi_\CD}{\partial\theta}\; . 
 \end{align}
 \label{eq:jac}
 \end{subequations}
The three derivatives sum to zero as the sum rule requires. 
Note the structure: the $\partial g^\CD$
piece is shared by $\Omega^\CD_\CQ$ and $\Omega^\CD_\CW$, while the
$\partial w^\Phi_\CD$ piece enters with opposite signs. 
The split is therefore the soft direction and the sum the stiff one. Recall here that the sum 
corresponds to the quantity that mimics ``dark energy'' in the energy budget. 
This is quantified in the next paragraph.\\

\noindent
\textit{Covariance and results, shown in Figure~\ref{fig:errors}}

We use the CPL constraints of Table~2 in \cite{mangangDESI} for the 
CMB data (\textit{Planck}+ACT\,DR6)+DESI\,DR2+DES-SN5YR, 
which reproduce the DESI collaboration's own values to within the last digit:
$w_0=-0.753\pm0.056$ (against DESI's $-0.752$), $w_a=-0.86^{+0.24}_{-0.21}$ ($\sigma_{w_a}=0.225$ symmetrized),
$\Omega_m=0.3192\pm0.0055$; we use DESI's central values
throughout and the errors of \cite{mangangDESI}.\footnote{In \cite{mangangDESI} the authors use the `Wang parametrization' \cite{wangDESI} that is a reparametrization of CPL  by ($w_0,w_{0.5}$); the equation of state  $w_{0.5}:=w(z{=}0.5)$ is listed in their Table~3 as $w_{a,\rm DE}$.}

The correlation $\rho:=\mathrm{corr}(w_0,w_a)$ is not quoted directly, but it can be
derived. The same work also reports the equation of state at
$z=0.5$, $w(z{=}0.5)=-1.039^{+0.035}_{-0.031}$, and for the CPL form $w(z)$ is a linear function of $(w_0,w_a)$ with $w(z)=w_0+w_a\,x$, $x=z/(1+z)$. Hence, 
$\sigma^2(w(z))=\sigma^2_{w_0}+x^2\sigma^2_{w_a}+2x\rho\,\sigma_{w_0}\sigma_{w_a}$,
and with $x=1/3$,
\begin{equation}
 \rho=\frac{\sigma^2\big(w(z{=}0.5)\big)-\sigma^2_{w_0}-\tfrac19\sigma^2_{w_a}}
 {\tfrac23\,\sigma_{w_0}\sigma_{w_a}}=-0.913 \; .
 \label{eq:rho}
 \end{equation}
The other two data combinations give $-0.980$ (CMB+DESI) and $-0.883$ (+Pantheon$+$). The asymmetry of
the quoted error bars moves $\rho$ between $-0.929$ and $-0.897$, i.e. the pivot
between $z_p=0.288$ and $0.301$. We adopt $\rho=-0.913$.

\begin{table*}\centering
\begin{tabular}{llll}
\toprule
 & $w^\Phi_\CD(\tilde a)$ & $f(z)$ & $w^\Phi_\CD(\infty)$ \\
\midrule
CPL & $w_0+w_a(1-\tilde a)$
    & $(1+z)^{3(1+w_0+w_a)}e^{-3w_az/(1+z)}$ & $w_0+w_a$\\[3pt]
BA  & $w_0+w_a\dfrac{1-\tilde a}{\tilde a^2+(1-\tilde a)^2}$
    & $(1+z)^{3(1+w_0)}\big(1+z^2\big)^{3w_a/2}$ & $w_0+w_a$\\[6pt]
JBP & $w_0+w_a\,\tilde a(1-\tilde a)$
    & $(1+z)^{3(1+w_0)}\exp\!\big[\tfrac32 w_a z^2/(1+z)^2\big]$ & $w_0$\\
\bottomrule
\end{tabular}
\captionsetup{width=0.98\linewidth, format=hang}
\caption{We consider three parametrizations to study the effect on adimensional backreaction functionals:
Barboza--Alcaniz (BA) \cite{BA}, unlike CPL, stays finite for all $\tilde a>0$, including the future
$\tilde a\to\infty$. Jassal--Bagla--Padmanabhan (JBP) \cite{JBP} confines the variation to
intermediate $z$, $w(0)=w(\infty)=w_0$.}
\label{tableIII}
\end{table*}

$\Omega_m^{\DT}$ is treated as uncorrelated (its $1.7\%$ error is subdominant for the
split). A $2.5\times10^4$-sample Gaussian Monte Carlo gives, at $z=0$,
\begin{align}
 \Omega_\mathcal{Q}^{\DT}=-0.214\pm0.029\ ,\qquad
 \Omega_\mathcal{W}^{\DT}=+0.895\pm0.029\ ,\nonumber\\
 \mathrm{corr}\big(\Omega^\CD_\CQ,\Omega^\CD_\CW\big)=-0.98 \; ,\qquad\qquad\qquad
\end{align}
in agreement with linear propagation through \eqref{eq:jac} to three digits, while for the sum,
\begin{equation}
 \sigma\big(\Omega_\mathcal{Q}^{\DT}+\Omega_\mathcal{W}^{\DT}\big)
 =\sigma\big(\Omega_m^{\DT}\big)=0.0055\; .
\end{equation}
We conclude that the total geometric budget that replaces the dark energy component is pinned five times better than the split between backreaction and curvature. The phantom-divide redshift is
$z_\times=0.405^{+0.076}_{-0.053}$. 
A minimum of $\sigma(\Omega^\CD_\CW)$ appears at the pivot redshift.

\section{Robustness against the choice of $w(\tilde a)$}
\label{robustness}

Equations~\eqref{eq:QWgeneral} and the $\Omega^\CD$'s follow solely from 
$w^\Phi_\CD=w(\tilde a)$ and the averaged energy constraint \eqref{averagehamilton}. 
Eliminating $f$ and $E$, we obtain the general expressions:
\begin{align}
 \Omega_\CQ^\CD(z)&=\tfrac14\big(1-\Omega_m^\CD(z)\big)\big(1+3w^\Phi_\CD(z)\big), \nonumber\\
 \Omega_\CW^\CD(z)&=\tfrac34\big(1-\Omega_m^\CD(z)\big)\big(1-w^\Phi_\CD(z)\big),
 \label{eq:universal}
\end{align}
i.e.\ $\big(\Omega_\CQ^\CD,\Omega_\CW^\CD\big)$ at redshift $z$ are
\textit{universal functions} of the two observables $\Omega_m^\CD(z)=\Omega_m^{\DT}(1+z)^3/E^2(z)$
and $w^\Phi_\CD(z)$ at the same $z$. The robustness of the reconstructed geometry traces back to 
the robustness of the pair $\big(E(z),w^\Phi_\CD (z)\big)$. Note that we could here substitute a binned or nonparametric
$w^\Phi_\CD (z)$ directly, and no parametrization would appear anywhere. This observation is investigated in
Appendix~\ref{nonparametric}.

We compare the three different parametrizations listed in Table~\ref{tableIII} and 
first fix the functional forms at $z=0$. We ask the following question: given one measurement of $\{w_0,w_a\}$, how much does the closure-inferred backreaction depend on the function or the shape of the continuation. This measures the parametrization, a best-fit of the functional forms to DESI along the lines of DESI's own fits in \cite{DESI} is provided thereafter.

\begin{table*}\centering
\begin{tabular}{@{}lccc|cc||ccc|cc@{}}
\toprule
 & \multicolumn{5}{c||}{$\Omega_\CQ^\CD$}
 & \multicolumn{5}{c}{$\Omega_\CW^\CD$}\\
$z$ & CPL & BA & JBP & $\Delta_{\rm BA}/\sigma$ & $\Delta_{\rm JBP}/\sigma$
    & CPL & BA & JBP & $\Delta_{\rm BA}/\sigma$ & $\Delta_{\rm JBP}/\sigma$\\
\midrule
$0$   & $-0.214$ & $-0.214$ & $-0.214$ & $0.0$ & $0.0$
      & $0.895$ & $0.895$ & $0.895$ & $0.0$ & $0.0$\\
$0.5$ & $-0.220$ & $-0.274$ & $-0.195$ & $-5.2$ & $+2.4$
      & $0.635$ & $0.665$ & $0.620$ & $+1.9$ & $-0.9$\\
$1$   & $-0.136$ & $-0.148$ & $-0.116$ & $-1.6$ & $+2.5$
      & $0.349$ & $0.303$ & $0.361$ & $-3.9$ & $+0.9$\\
$2$   & $-0.041$ & $-0.024$ & $-0.042$ & $+7.6$ & $-0.3$
      & $0.097$ & $0.046$ & $0.134$ & $-5.3$ & $+3.9$\\
$3$   & $-0.014$ & $-0.005$ & $-0.019$ & $+5.1$ & $-2.6$
      & $0.032$ & $0.010$ & $0.062$ & $-3.6$ & $+5.0$\\
\bottomrule
\end{tabular}
\captionsetup{width=0.7\linewidth, format=hang}
\caption{Functional-form dependence at fixed $(w_0,w_a,\Omega_m^{\DT})$, compared with
the propagated DESI uncertainty ($\Delta_{\rm X}:=\Omega^{\rm X}-\Omega^{\rm CPL}$).}
\label{tableIV}
\end{table*}

\begin{figure*}\centering
\includegraphics[width=0.7\textwidth]{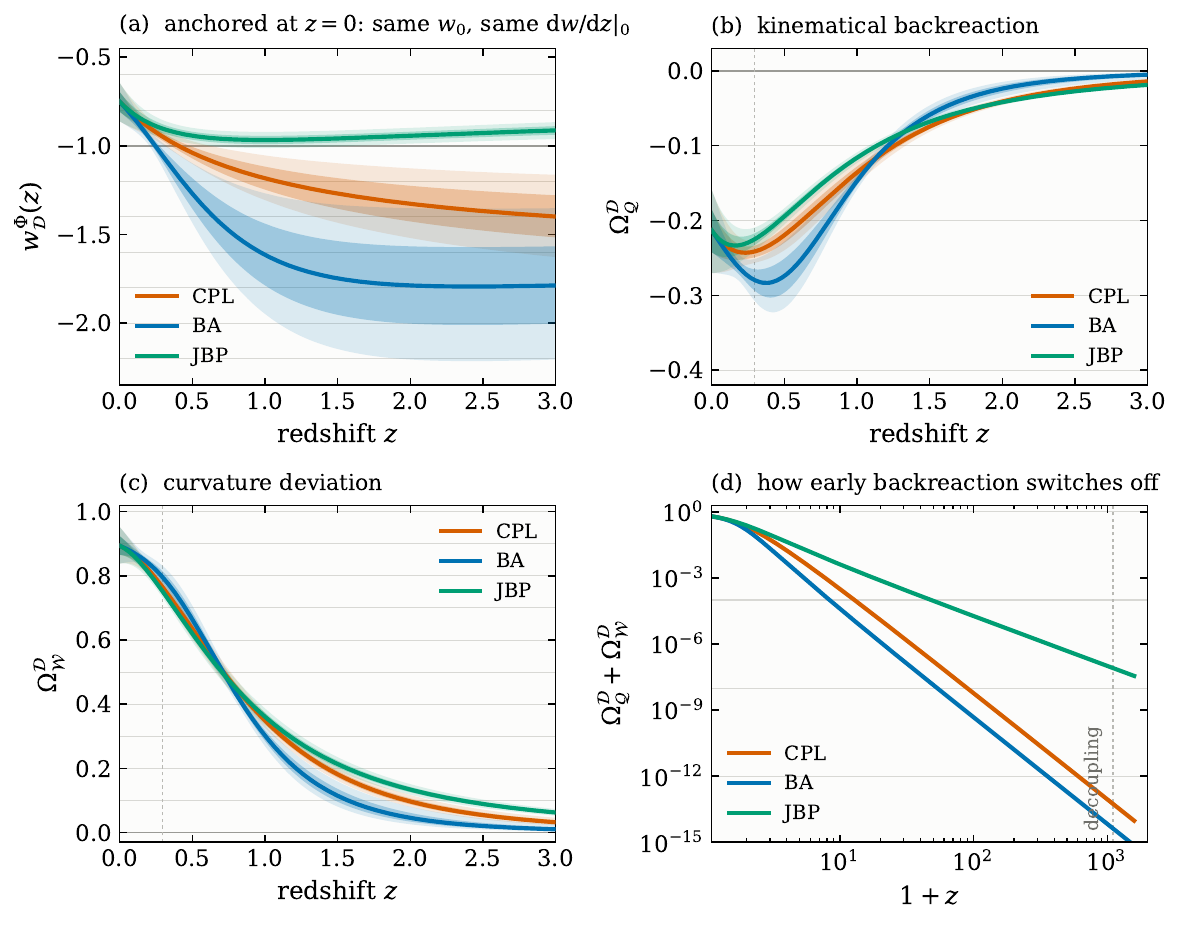}
\caption{The three forms with the same Taylor data at $z=0$ ($w_0, w_a, \Omega_m^\DT)
= (-0.752, -0.86, 0.3191)$ imposed on all three, fed through the same closure.
(a) the equations of state; the grey horizontal line marks $w^\Phi_\CD=-1$. (b),(c) the $\Omega^\CD$'s, with the
propagated DESI $68\%$ band of Fig.~\ref{fig:errors} to appreciate the relevant scale. (d) the
early-time behavior of the total geometrical budget. We notice that, although the redshift dependence of the equations of state looks wildly different if just continued from $z=0$, the qualitative redshift evolutions of the energies in backreaction and curvature deviation are fairly robust.}
\label{fig:params}
\end{figure*}

\begin{table*}
\begin{center}
\begin{tabular}{lccccc}
\toprule
 & $w_0$ & $w_a$ & $\Omega^\DT_m$ & $\rho(w_0,w_a)$ & $\chi^2/\mathrm{dof}$\\
\midrule
CPL & $-0.764\pm0.093$ & $-0.734\pm0.542$ & $0.3218\pm0.0148$ & $-0.90$ & $8.00/10$\\
BA  & $-0.790\pm0.074$ & $-0.369\pm0.260$ & $0.3225\pm0.0146$ & $-0.83$ & $7.71/10$\\
JBP & $-0.737\pm0.123$ & $-1.042\pm0.933$ & $0.3126\pm0.0139$ & $-0.94$ & $8.48/10$\\
\midrule
$\Lambda$CDM & $-1$ & $0$ & $0.3124$ & --- & $18.64/12$\\
\bottomrule
\end{tabular}
\end{center}
\captionsetup{width=0.95\linewidth, format=hang}
\caption{Three genuine DESI DR2 fits pushed through the closure. 
The CPL row stays within $0.23\sigma$ of the published DESI$+$CMB$+$DESY5 values
$(-0.752,-0.86,0.3191)$; (errors are larger than DESI's, which include SNe; here the differences between the forms matter, and those are robust).}
\label{tableV}
\end{table*}

\begin{figure*}\centering
\includegraphics[width=\textwidth]{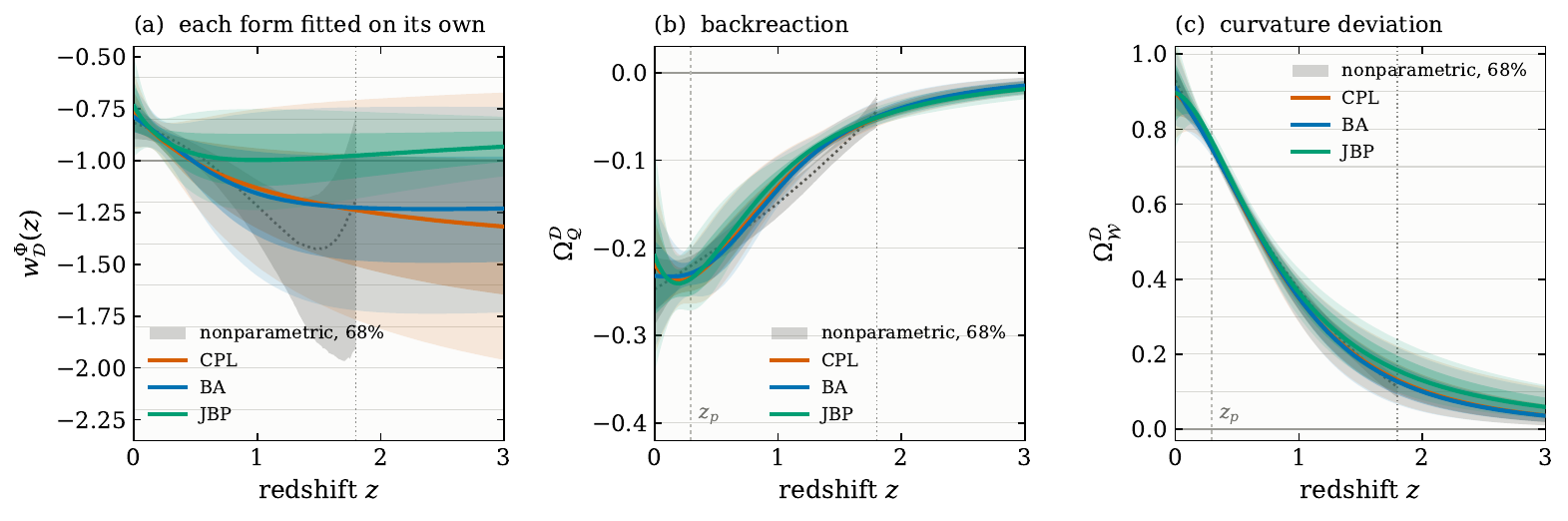}
\caption{Each form fitted to DESI DR2 BAO with the
CMB calibration $c/(H_\DT r_d)=30.467\pm0.255$ as a prior, free $(w_0,w_a,\Omega^\DT_m,\kappa)$; errors are from the Gauss-Newton covariance at the maximum, which is a Gaussian approximation to the posterior and is why the bands are symmetric (unlike DESI's bands in \cite{DESI} (Fig.~4), since their $w_0$--$w_a$ posterior is banana-shaped). 
(a) the three fitted $w^\Phi_\CD(z)$; (b) the resulting $\Omega_\CQ^\CD$, fanning out at $z=0$ and pinching at $z_p$; (c) the resulting $\Omega_\CW^\CD$. The $68\%$ confidence level of the nonparametric closure of Appendix~\ref{nonparametric} is included.}
\label{fig:refit}
\end{figure*}

\begin{table}
\begin{center}
\begin{tabular}{lcccccc}
\toprule
$z$ & $0$ & $0.22$ & $0.294$ & $0.32$ & $0.5$ & $1$\\
\midrule
anchored spread in $\Omega^\CD_\CQ$ & $0$ & $0.037$ & $0.054$ & $0.059$ & $0.079$ & $0.032$\\
\hspace{3mm} in units of $\sigma_{\rm stat}$ & $0$ & $4.1$ & $6.9$ & $7.5$ & $7.7$ & $4.1$\\
refitted spread in $\Omega^\CD_\CQ$ & $0.023$ & $0.009$ & $0.006$ & $0.005$ & $0.006$ & $0.013$\\
refitted spread in $\Omega^\CD_\CW$ & $0.014$ & $0.021$ & $0.019$ & $0.018$ & $0.009$ & $0.013$\\
\bottomrule
\end{tabular}
\caption{Comparison of the spreads if different parametrizations are anchored at $z=0$ (monotonically 
increasing spread out to $z\simeq0.5$), with the refitted case relevant for the comparison with data in Figure~\ref{fig:refit}.}
\label{tableVI}
\end{center}
\end{table}

We vary the functional forms at fixed $(w_0,w_a)$, i.e.\ fixing $w(0)=w_0$ and
$\mathrm{d}w/\mathrm{d}z|_0=w_a$.
Since $f(1)=1$ and $w^\Phi_\CD(0)=w_0$ for all of them, \eqref{eq:universal} gives
\begin{equation}
 \Omega^{\DT}_\CQ=\frac{\big(1-\Omega_m^{\DT}\big)\big(1+3w_0\big)}{4}, \ \ 
 \Omega^{\DT}_\CW=\frac{3\big(1-\Omega_m^{\DT}\big)\big(1-w_0\big)}{4} ,
 \label{eq:anchors}
\end{equation}
so at fixed $(w_0,\Omega_m^{\DT})$ the present-day split is identical for every
form: $-0.214$ and $+0.895$. The first derivatives agree too:
$\mathrm{d}\Omega_\CQ^\CD/\mathrm{d}z|_0=-0.28526$,
$\mathrm{d}\Omega_\CW^\CD/\mathrm{d}z|_0=-0.20492$ for all three of the parametrizations CPL, BA, and JBP.
The form first enters at $O(z^2)$: $\mathrm{d}^2\Omega_\CQ^\CD/\mathrm{d}z^2|_0$ is $+1.66$ (CPL),
$-0.10$ (BA), $+2.54$ (JBP). Further out the differences reach high significance, 
\textit{cf.} Table~\ref{tableIV} and Fig.~\ref{fig:params}.
At $z=0$ the three reconstructions trivially agree, because \eqref{eq:anchors} is evaluated at
$w_0$, which all three share by construction. This is an artefact of the anchoring,
not a measurement. The shape systematic is set to zero where
$w^\Phi_\CD$ is worst measured. It reappears immediately, reaching $0.054$ in $\Omega^\CD_\CQ$ at the pivot 
and $0.079$ at $z=0.5$. Of course, the systematic to be quoted in the comparison with observed data is the one obtained when each form is allowed its own fit (Tables~\ref{tableV}, \ref{tableVI} and Figure~\ref{fig:refit}). 

As a practical conclusion, if different parametrizations have to be compared with the aim to match the observations,
the backreaction reconstruction could be quoted at the pivot, not at $z=0$.
This is the backreaction analogue of the standard advice to quote $w_p$ rather than
$w_0$, and it follows from \eqref{eq:universal}: the reconstruction is only as
good as $w^\Phi_\CD(z)$ at that $z$.

Anchoring the fit to $z=0$ two structural differences arise. 
First, concerning the \textit{phantom crossing}: JBP crosses $w=-1$ only if $|w_a|>4(1+w_0)=0.99$, since $\max_{\tilde a}\tilde a(1-\tilde a)=1/4$, with $w_a=-0.86$ it does \textit{not} cross ($\min w^\Phi_\CD=-0.967$) and the morphon is a
standard scalar throughout. BA crosses at $z_\times=0.246$ and overshoots the common
asymptote, reaching $w^\Phi_\CD=-1.790$ at $z=2.41$. 
Second, for the \textit{Early Universe}, we have at decoupling that
$\Omega_\CQ^\CD+\Omega_\CW^\CD$ is $4\times10^{-15}$ (BA),
$5\times10^{-14}$ (CPL), $8\times10^{-8}$ (JBP), the spread coming from
$w^\Phi_\CD(\infty)$ being $w_0+w_a$ or $w_0$. Thus, backreaction is negligible at decoupling in
all of them. Of course, current data lack the sensitivity to distinguish different parametrizations beyond $z \approx 2$.

Now we turn to fitting the different parametrizations to DESI data in the spirit of DESI's own comparison of different parametrizations. 
Fixing $(w_0,w_a)$ is a controlled test of shape, but not what happens in practice: if a
different form is refitted, $w_0$ moves. Table~\ref{tableV} and Figure~\ref{fig:refit} 
fit each of the three forms independently to the DESI DR2 BAO distances with the CMB calibration 
of $c/(H_\DT r_d)$ as a prior.
Each curve is then a legitimate description of the observations, and the spread is the
real parametrization systematic. 
For the error budget the refitted values are relevant:
\begin{align*}
\Omega^\CD_\CQ(z_p)&\simeq-0.242 \;\pm\; 0.008\,({\rm stat.})\;\pm\;0.006\;({\rm syst.})\; ,\\
\Omega^\CD_\CW(z_p)&\simeq+0.766 \;\pm\; 0.012 \;({\rm stat.})\;\pm\;0.019\;({\rm syst.})\; .
\end{align*}

We conclude that again the backreaction and curvature deviation evolution is fairly robust, but uncertainties fan out
at the present day and after the pivot. 

\section{Nonparametric closure}
\label{nonparametric}

Appendix~\ref{robustness} stated a universal map showing that the closure only needs $\Omega^\CD_m(z)$ and $w^\Phi_\CD(z)$. In fact, we can go a step further and eliminate $w^\Phi_\CD$ as well. 
As indicated there, we may even drop parametrizations altogether and feed a binned or nonparametric reconstruction of $w^\Phi_\CD (z)$ (also published by the DESI team \cite{DESI}) into \eqref{eq:universal}. 
Then $\CQ_\CD$ and $\CW_\CD$ inherit the data's own error
structure instead of a functional-form prior and, according to what has been found in Appendix~\ref{robustness}, the quantities to quote are those at the pivot, where both error budgets are minimal.
The closure can be written as a local functional of the expansion history alone, with no $w^\Phi_\CD$ and
no dark-energy variable anywhere. Only $E(z)$ is needed.

With $\Lambda=k_{\CD_i}=0$, the averaged
Raychaudhuri equation \eqref{averageraychaudhuri} and the averaged energy constraint \eqref{averagehamilton}
give, using $\mathrm{d}/\mathrm{d}t=-(1+z)H_\CD\,\mathrm{d}/\mathrm{d}z$ and
$\ddot a_\CD/a_\CD=H_\CD^2-(1+z)H_\CD\,\mathrm{d}H_\CD/\mathrm{d}z$, 
\begin{subequations}
\label{Edependence}
\begin{align}
\frac{\CQ_\CD(z)}{H^2_{\DT}} &= 3E^2(z)-\frac{3}{2}\,(1+z)\,\frac{\mathrm{d}E^2}{\mathrm{d}z}
   +\frac{3}{2}\,\Omega^{\DT}_m(1+z)^3 \; ,
\label{eq:Qe}\\
\frac{\CW_\CD(z)}{H^2_{\DT}} &= 6\,\Omega^{\DT}_m(1+z)^3-6E^2(z)-\frac{\CQ_\CD(z)}{H^2_{\DT}}\; ,
   \label{eq:We}
\end{align}
\end{subequations}
with $E:=H_\CD/H_{\DT}$. Only the observed expansion history and the domain mass
enter. There is no equation of state and no dark-energy density.

Recall that a cosmological constant is the statement that both geometrical terms are constant,
\begin{subequations}
\begin{align}
\CQ_\CD^\Lambda&=3\big(1-\Omega^{\DT}_m\big)H^2_{\DT}=2.043\,H^2_{\DT}\; , \\
\CW_\CD^\Lambda&=-9\big(1-\Omega^{\DT}_m\big)H^2_{\DT}=-6.128\,H^2_{\DT}\; ,
\end{align}
\end{subequations}
for $\Omega^{\DT}_m=0.3191$ stating a null hypothesis that needs no parametrization to test. 
Then, observe that \eqref{Edependence} is local in $z$: $\CQ_\CD$ and $\CW_\CD$ at a given redshift are
fixed by $E^2$ and its first derivative. 
The reconstruction is therefore limited only by how well $\mathrm{d}E^2/\mathrm{d}z$ is measured.

\begin{table*}\centering
\begin{tabular}{@{}lcc|cc|cc@{}}
\toprule
 & \multicolumn{2}{c|}{$\Omega^\CD_\CQ$}
 & \multicolumn{2}{c|}{$\Omega^\CD_\CW$}
 & \multicolumn{2}{c}{$\CQ_\CD/H^2_{\DT}$}\\
$z$ & non-param. & CPL & non-param. & CPL & non-param. & CPL\\
\midrule
$0$     & $-0.249\pm0.038$ & $-0.214\pm0.029$ & $+0.931\pm0.038$ & $+0.895\pm0.029$ & $1.50\pm0.23$ & $1.28$\\
$0.294$ & $-0.221\pm0.012$ & $-0.242\pm0.008$ & $+0.747\pm0.009$ & $+0.766\pm0.012$ & $1.92\pm0.07$ & $2.11$\\
$0.5$   & $-0.201\pm0.012$ & $-0.220\pm0.010$ & $+0.623\pm0.018$ & $+0.635\pm0.016$ & $2.24\pm0.14$ & $2.43$\\
$1$     & $-0.148\pm0.015$ & $-0.136\pm0.008$ & $+0.373\pm0.020$ & $+0.349\pm0.012$ & $2.92\pm0.28$ & $2.65$\\
$1.5$   & $-0.086\pm0.011$ & $-0.075\pm0.004$ & $+0.192\pm0.022$ & $+0.181\pm0.011$ & $2.88\pm0.34$ & $2.50$\\
$2$     & $-0.016\pm0.028$ & $-0.041\pm0.002$ & $+0.073\pm0.032$ & $+0.097\pm0.009$ & $0.9\pm1.5$   & $2.26$\\
\bottomrule
\end{tabular}
\captionsetup{width=0.8\linewidth, format=hang}
\caption{The closure evaluated non-parametrically from the DESI DR2 BAO expansion
history, against the CPL closure of the main text. Errors are $68\%$.}
\label{tableVII}
\end{table*}

\begin{figure*}\centering
\includegraphics[width=0.77\textwidth]{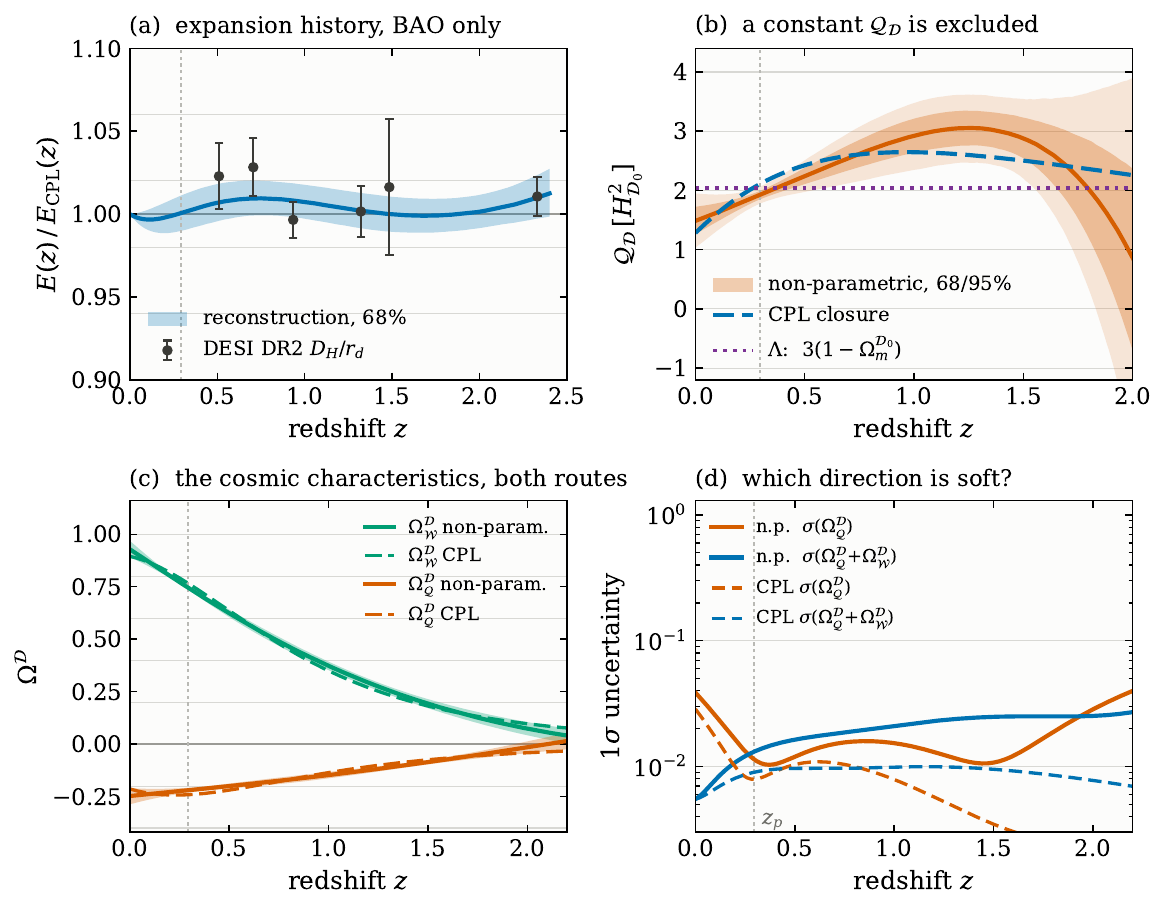}
\captionsetup{width=0.9\linewidth, format=hang}
\caption{(a) the BAO-only reconstruction of $E(z)$, divided by the CPL best fit, with
the $D_H/r_d$ points; (b) $\CQ_\CD$ with the $\Lambda$ benchmark; (c) both routes to the
cosmic characteristics; (d) the uncertainty on the split and on the sum, for the two routes.
The dotted vertical line is the pivot $z_p=0.294$.}
\label{fig:nonparam}
\end{figure*}

\subsection{Data and method}

We refer to Table~IV in \cite{DESI-II} for notations and for the DESI DR2 BAO tracers with both $D_M/r_d$ and $D_H/r_d$ ($z_{\rm eff}=0.510$, $0.706$, $0.934$, $1.321$, $1.484$, $2.330$), including the
quoted within-tracer correlations $r(D_M,D_H)\simeq-0.40$ to $-0.50$, plus the Bright Galaxy Survey (BGS)
$D_V/r_d$ at $z=0.295$. 
$D_H/r_d=c/(H_\CD r_d)$ measures $E(z)$ directly; $D_M/r_d$ supplies the integral
constraint that gives leverage below the lowest $D_H$ point, which matters because the
pivot $z_p=0.294$ lies there. 

We expand the \textit{expansion history}, not the dark energy,
\begin{equation}
\label{expansionhistory}
 \ln E^2(z)=\sum_{k=1}^{K} c_k\,x^k,\qquad x:=\ln(1+z)\; ,
\end{equation}
which enforces $E(0)=1$ and has no constant term. The normalization
$\kappa:=c/(H_{\DT}r_d)$ is a free parameter with the Gaussian prior
$30.467\pm0.255$ implied by $H_0=66.73\pm0.55$ and $r_d=147.46\pm0.21$ of
\cite{mangangDESI}, i.e.\ the same calibration already used in Appendix~\ref{uncertainties}. 
$\Omega^{\DT}_m=0.3191\pm0.0055$ is the single external input (beyond the $\kappa$-calibration); the domain mass is not a BAO observable. $K=3$ is adopted; $\chi^2=6.9$ for $10$ degrees of freedom, and $K=4$
does not improve the fit ($\Delta\chi^2=0.2$ for one extra parameter). Uncertainties
come from $4000$ Monte-Carlo realizations of the full data covariance plus the $\kappa$
and $\Omega^{\DT}_m$ priors, refitting each time. Table~\ref{tableVII} and Figure~\ref{fig:nonparam} 
show the results discussed below.

\subsection{Results}
\label{sec:res}

Applied to the DESI DR2 BAO distances this yields a
reconstruction of $\CQ_\CD$ and $\CW_\CD$ that allows for two main conclusions and a technical remark, 
(i) it confirms the CPL closure to within $1.5\sigma$ at every redshift, and $z_\times$ also differs by $1.5\sigma$,  
(ii) it disfavours a constant $\CQ_\CD$, i.e. a cosmological constant without any dark-energy parametrization assumed, and (iii) it reproduces the $z=0$ stiffness of the sum $\Omega_\CQ^\CD+\Omega^\CD_\CW$ found in Appendix~\ref{uncertainties}. 

\begin{itemize}
\item[(i)] \textbf{The CPL closure is confirmed.} The two reconstructions agree
within $1.5\sigma$ at every redshift, when both errors are combined (the discrepancies are at most $1.46\sigma$
with the errors combined and $1.75\sigma$ against the nonparametric error alone), 
and the qualitative structure of the main text
survives: $\CQ_\CD>0$ and $\CW_\CD<0$ throughout, $\CQ_\CD$ rising to $2.92\pm0.28$ at $z=1$ nonparametrically against $2.65$ for CPL. 
The price is a $1.5$--$2\times$ larger uncertainty at
the pivot and a complete loss of constraint beyond $z\simeq1.8$. The central values
shift by $0.011$--$0.035$ in $\Omega^\CD_\CQ$, which is larger than the
$0.006$ parametrization systematic estimated in Appendix~\ref{robustness},  
understating the true functional-form error.

\item[(ii)] \textbf{A cosmological constant is disfavoured without parametrizing the
equation of state.} We compare the reconstructed $\CQ_\CD(z)$ with the constant $3(1-\Omega^{\DT}_m)$:
\begin{center}\small
\begin{tabular}{lcccccc}
\toprule
$\textrm{redshift} \ z$ & $0$ & $0.294$ & $0.5$ & $1$ & $1.5$ & $2$\\
$(\CQ_\CD-\CQ_\CD^\Lambda)/\sigma$ & $-2.4$ & $-1.7$ & $+1.5$ & $+3.1$ & $+2.5$ & $-0.8$\\
\bottomrule
\end{tabular}
\end{center}
$\sigma$ is the error of the difference, which propagates
$\Omega_m^{\DT}$ through both $\CQ_\CD$ and $\CQ^{\Lambda}_{\CD}$.
The reconstruction is below the $\Lambda$ value at low redshift and above it at
$z\simeq1$, an evolving $\CQ_\CD$, not a constant one. This is the backreaction
counterpart of DESI's evidence for evolving dark energy, obtained here with no
assumption about $w^\Phi_\CD$. Globally, the reconstruction improves on
a constant $\CQ_\CD$ by $\Delta\chi^2=11.7$ for two extra parameters ($18.64/12$ of
Table~\ref{tableV} against $6.91/10$), i.e. at $2.98\sigma$. Note, however, that the pointwise
entries of the above table do not carry this significance, since all six entries of the above table are strongly correlated.
For the $K$-dependence, taking instead of $K=3$, $K=2$ and $K=4$ gives 
$+4.8$ and $+2.0$ instead of $+3.1$. 
The redshift dependence of $\CQ_\CD$, not its amplitude, is what the data constrain.

\item[(iii)] \textbf{The degeneracy direction} is a property of the prior, not of the physics. 
Appendix~\ref{uncertainties} finds the sum $\Omega^\CD_\CQ+\Omega^\CD_\CW$
pinned five times better than the split. That statement is specific to $z=0$, where $E(0)=1$ holds by construction, 
so the sum reduces to $\sigma(\Omega^{\DT}_m)$; already at the pivot the CPL ratio is
$\sigma(\mathrm{sum})/\sigma(\Omega^\CD_\CQ)=1.14$, not $1/5$. The nonparametric closure, which
enforces $E(0)=1$ through \eqref{expansionhistory} and therefore shares the $z=0$ statement
($\sigma(\mathrm{sum})=\sigma(\Omega^{\DT}_m) = 0.0055$, ratio $0.14$), behaves the same way:

\smallskip
\begin{center}\small
\begin{tabular}{lccccc}
\toprule
redshift $z$ & 0 & 0.294 & 0.5 & 1 & 1.5\\
CPL: $\sigma(\mathrm{sum})/\sigma(\Omega^\CD_\CQ)$ & 0.19 & 1.14 & 0.93 & 1.25 & 2.74\\
nonparam.: $\sigma(\mathrm{sum})/\sigma(\Omega^\CD_\CQ)$ & 0.14 & 1.11 & 1.37 & 1.40 & 2.26\\
\bottomrule
\end{tabular}
\end{center}

\end{itemize}

\section{Bounds on the morphon kinetic term}
\label{koksbang}

A reconstruction of the angular diameter distance and the effective expansion rate
together with their redshift derivatives from supernova and BAO data without
assuming an FLRW model has been given by Koksbang \cite{koksbang}, based on recent work by Heinesen and Clifton \cite{tests}. She derives bounds on the backreaction combination
$\Omega^\CD_\CR+3\,\Omega^\CD_\CQ$ over the redshift range $0.5<z<2.25$.  
The method is close kin to the nonparametric closure of Appendix~\ref{nonparametric}, a derivative of a reconstructed
expansion history, and it is the first direct bound of its kind.

With $k_{\CD_i}=0$, so that $\CW_\CD=\av{\CR}$, the combination she constrains is our morphon kinetic term \eqref{morphoncorrespondence}, also written in terms of the deceleration functional \eqref{eq:cosmo}:
\begin{align}
{\mathcal K}^\CD&:=\Omega^\CD_\CW+3\,\Omega^\CD_\CQ =-\frac{\CW_\CD+3\CQ_\CD}{6H_\CD^2}= 
1+q_\CD-\tfrac{3}{2}\,\Omega^\CD_m (z)\nonumber\\
&=\frac{4\pi G\,\epsilon\dot\Phi^2_\CD}{H_\CD^2}=\frac{3}{2}\,
\frac{\big(1+w^\Phi_\CD\big)\big(1-\Omega^{\DT}_m\big)f}{E^2} \, .
\label{eq:kok}
\end{align}
It therefore vanishes iff $w^\Phi_\CD=-1$, so Koksbang is measuring the deviation from the phantom divide characterized by $q_{\CD}=\frac32\Omega^\CD_m-1$, ${\dot H}_{\CD}=-4\pi G\langle\varrho\rangle_{\CD}$, without using a  parametrization of $w^\Phi_\CD$ and without a FLRW template. 

Our closure makes a sharp prediction
on this quantity: from equations \eqref{eq:Qe}, \eqref{eq:We} we obtain:
\begin{subequations}
\begin{align}
\CW_\CD +3\CQ_\CD&=\Big[9\,\Omega_m^{\DT}(1+z)^3-3(1+z)\frac{\mathrm{d}E^2}{\mathrm{d}z}\Big]H^2_{\DT}\; , \\
{\mathcal K}^\CD &=\frac{1+z}{2E^2}\,\frac{\mathrm{d}}{\mathrm{d}z}\Big[E^2-\Omega_m^{\DT}(1+z)^3\Big]\; .
\end{align}
\end{subequations}
The phantom divide $z_\times$ is the stationary point of $E^2-\Omega_m^{\DT}(1+z)^3$, evaluated in Table~\ref{tableVIII} to be $$z_\times^{\rm nonparam.} = 0.58^{+0.16}_{-0.12}\; ,$$ where 
$\text{98\,\% of realizations cross in \{0.1<z<2\}}$, i.e. with the functional basis changed from $w(\at)$ to
$\ln E^2$ (with no dark-energy variable), the phantom crossing is still there without being a property of parametrizations.

\begin{table}[h]
\begin{center}\small
\begin{tabular}{@{}lrrrrrr@{}}
\toprule
$z$ & $0$ & $0.405$ & $0.5$ & $1.0$ & $1.5$ & $2$\\
\midrule
$\CK^\CD$, CPL              & $+0.253$ & $0.000$ & $-0.024$ & $-0.058$ & $-0.043$ & $-0.027$\\
$\sigma$                & $0.057$  & $0.019$ & $0.020$  & $0.019$  & $0.011$  & $0.005$\\
$s$                     & $4.4$    & ---     & $1.2$    & $3.0$    & $3.9$    & $5.0$\\[2pt]
$\CK^\CD$, nonpar.       & $+0.186$ & $+0.052$& $+0.023$ & $-0.071$ & $-0.066$ & $+0.030$\\
$\sigma$                & $0.078$  & $0.031$ & $0.032$  & $0.044$  & $0.040$  & $0.069$\\
$s$                     & $2.4$    & $1.7$   & $0.7$    & $1.6$    & $1.7$    & $0.4$\\
\bottomrule
\end{tabular}
\end{center}
\caption{Closure prediction for $\CK^\CD(z)$. Upper rows: the CPL closure with
the DESI DR2 covariance of Appendix~\ref{uncertainties} propagated; the significance
$s:=|{\rm value}|/\sigma$ is conditional on the two-parameter form, and is
undefined at $z_\times$ where $\CK^\CD$ vanishes. Lower rows: the
parametrization-free $K=3$ reconstruction of Appendix~\ref{nonparametric}, where the errors are
those of the data.}
\label{tableVIII}
\end{table}

Several features make this a good test.
Under the CPL closure it is sign-definite, negative throughout $0.5<z<2$ (\textit{cf.} Table~\ref{tableVIII}). It vanishes at $z_\times$, just below the analyzed redshift range. Evaluated instead on the non-parametric expansion history, the
central values agree within about $1\sigma$ everywhere, but the significance per
redshift drops to $\lesssim1.7\sigma$, and at $z=0.5$ the sign is not established 
(the non-parametric divide lies at $z_\times\simeq0.58$, so
$z=0.5$ falls inside the crossing region). The sign of $\CK^\CD$ is therefore a prediction to
be tested, not yet a measurement.

\begin{figure}[h]\centering
\includegraphics[width=0.45\textwidth]{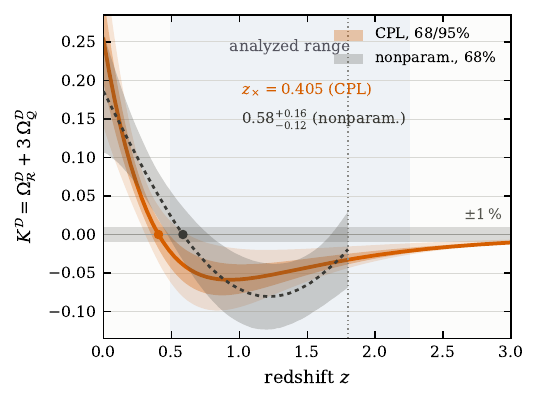}
\caption{The closure's prediction for the combination ${\mathcal K}^\CD$ with the DESI DR2
uncertainties propagated (CPL: Appendix~\ref{uncertainties}). Overlaid is the non-parametric
$68\,\%$ band, cut at $z=1.8$. Shaded band: the analyzed redshift range by Koksbang \cite{koksbang}. 
Grey horizontal band: $\pm1\,\%$, the percent level.}
\label{fig:koksbangbound}
\end{figure}

Figure~\ref{fig:koksbangbound} shows that extending the method below $z=0.5$ would be highly desirable,
especially since the closure-predicted signal becomes an order of magnitude larger than at $z=0.5$. 
Against her statement that the data cannot yet separate zero from percent level and would not allow to exclude 
substantial backreaction, 
the closure predicts a few percent: the test is at, and just beyond, the current frontier.

A complementary prediction is obtained by using Eq.~\eqref{eq:syksy} of Appendix~\ref{distances}. 
This equation is equivalent to
$E\,(E\,D')'=\CK^\CD E^2 D/(1+z)^2$ for $D=(1+z)\bar D_A H_\DT$.\footnote{The coefficient appearing here is 
$-\mathcal{K}_{\rm BR}$, with the effective curvature function of Heinesen and Clifton \cite{tests}, $\mathcal{K}_{\rm BR}=(\av{\CR}+3\CQ_\CD) / (6\,(1+z)^2)$. For $\CQ_\CD = 0$ one has $\av{\CR} =6 k (1+z)^2$, so $\mathcal{K}_{\rm BR}$ 
reduces to the FLRW curvature constant $k$.} 
Integrating once, we obtain:
\begin{align}
\label{CBL}
\mathcal{O}^\CD (z):=\frac{(E\,D')^2-1}{D^2}=\frac{1}{D^2}\int_0^{z}\frac{\CK^\CD E^2}{(1+z')^2}\;\mathrm{d}\!\left(D^2\right),
\end{align}
where $\CK^\CD E^2/(1+z)^2=4\pi G\,\epsilon\dot\Phi_\CD^2\,\tilde a^2/H_\DT^2$. This shows that 
the Clarkson--Bassett--Lu function\footnote{We normalize (D3) as in [28]: with the distance measured in units of the Hubble radius $c/H_\DT$ and $E:=H_{\CD}/H_\DT$, $[(E D')^2-1]/D^2$ is identical to
$[(H_{\CD}\mathfrak{D}')^2-1]/(H_\DT\mathfrak{D})^2$ with $\mathfrak{D}$ the
dimensionful comoving (transverse) distance, $\mathfrak{D}=(1+z){\bar D}_A$, and $'=\mathrm{d}/\mathrm{d}z$, which is the function of \cite{curvaturetest} and returns the constant
$\Omega_k$ in any FLRW model. Ref.~\cite{tests} writes the same object without this
normalization, $\mathcal{O}=[H^2(\mathfrak{D}')^2-1]/\mathfrak{D}^2$, which carries the
dimensions of a curvature and reduces to $-k$; the two differ by the constant factor
$H^2_\DT$ and carry the same information. We keep the dimensionless form, first
because all characteristics here are measured in units of $H^2_\DT$, so that
$\mathcal{O}^\CD(0)=\CK^\CD(0)$ is an identity between numbers, and second because a curvature-valued
diagnostic would reintroduce the normalization of $a_\CD$, which by \eqref{eq:norm}, \eqref{aD0dropped} has dropped
out of every observable. We write $\mathcal{O}^\CD(z)$ rather than $\Omega_k(z)$ because here the
function is a redshift-dependent diagnostic, not to be confused with a density parameter of the energy
budget. We have:
$$
\mathcal{O}^\CD=\frac{(E D')^2-1}{D^2}=\frac{\bigl(H_{\CD}\mathfrak{D}'\bigr)^2-1}{\bigl(H_\DT\mathfrak{D}\bigr)^2}
=\Omega_k(z) \ \text{of \cite{curvaturetest}}=\frac{\mathcal{O}^{\cite{tests}}}{H^2_\DT}\;.
$$} \cite{curvaturetest}, \cite{tests} is the $D^2$-weighted line-of-sight mean of the morphon kinetic term times $\tilde a^2$. For the baseline closure we obtain the values of Table~\ref{tableIX} with $\mathcal{O}^\CD (0)=\CK^\CD(0)=\tfrac32(1+w_0)(1-\Omega^\DT_m)$ and a zero at $z_0=0.773$.

\begin{table}[h]
\begin{center}\small
\begin{tabular}{@{}lrrrrrrr@{}}
\toprule
$z$ & $0$ & $0.2$ & $0.405$ & $0.5$ & $1.0$ & $1.5$\\ 
\midrule
$\mathcal{O}^\CD$, CPL          & $+0.253$ & $+0.131$ & $+0.059$ & $+0.038$ & $-0.015$ & $-0.027$\\
$\sigma$            & $0.057$  & $0.032$  & $0.018$  & $0.016$  & $0.014$  & $0.013$\\
$s$                 & $4.4$    & $4.1$    & $3.2$    & $2.4$    & $1.1$    & $2.1$\\[2pt]
$\mathcal{O}^\CD$, nonpar.   & $+0.186$ & $+0.131$ & $+0.087$ & $+0.069$ & $+0.004$ & $-0.026$\\
$\sigma$            & $0.078$  & $0.050$  & $0.034$  & $0.030$  & $0.027$  & $0.029$\\
$s$                 & $2.4$    & $2.6$    & $2.5$    & $2.3$    & $0.2$    & $0.9$\\
\bottomrule
\end{tabular}
\end{center}
\caption{Closure prediction for the curvature-consistency function $\mathcal{O}^\CD (z)$ of
Eq.~\eqref{CBL}.  
Upper rows: the CPL closure with the DESI DR2 covariance of Appendix~\ref{uncertainties}; lower rows: the
parametrization-free $K=3$ reconstruction of Appendix~\ref{nonparametric}, which is constrained to
$z\lesssim1.8$. $s:=|{\rm value}|/\sigma$. Every spatially flat FLRW model has
$\mathcal{O}^\CD = 0$, whatever the dark-energy equation of state.}
\label{tableIX}
\end{table}

\begin{figure*}\centering
\includegraphics[width=0.85\textwidth]{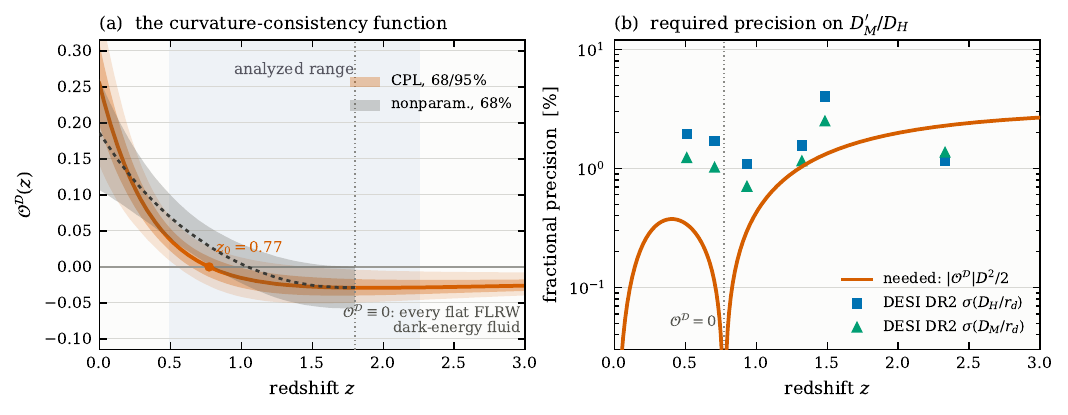}
\caption{Curvature consistency test using the Clarkson--Bassett--Lu function \cite{curvaturetest}, \cite{tests}.
(a) the prediction for $\mathcal{O}^\CD (z)$ in the format of Figure~\ref{fig:koksbangbound}. The CPL propagation
of the DESI DR2 covariance ($68/95\,\%$), the parametrization-free $68\,\%$ band, cut
at $z=1.8$, and Koksbang's analyzed range shaded; the CPL prediction vanishes at
$z_0=0.77$. 
(b) this panel shows what it would take for observational discrimination: the
fractional departure $\delta=|\mathcal{O}^\CD| D^2/2$ of $D_M'/D_H$ from the FLRW identity implied
by \eqref{CBL2}, against the DESI DR2 per-tracer precision of $D_H/r_d$ and $D_M/r_d$.}
\label{fig:curvaturetest}
\end{figure*}

\begin{table*}
\begin{center}\footnotesize
\begin{tabular}{@{}lrrrrrrr@{}}
\toprule
$z$ & $0.2$ & $0.405$ & $0.5$ & $1.0$ & $1.5$ & $2.0$ & $2.33$\\
\midrule
$\mathcal{O}^\CD$ (CPL closure) & $+0.131$ & $+0.059$ & $+0.038$ & $-0.015$ & $-0.027$ & $-0.029$ & $-0.028$\\
$D=D_MH_\DT/c$   & $0.187$ & $0.357$ & $0.429$ & $0.750$ & $0.991$ & $1.175$ & $1.274$\\
needed on $D_M'/D_H$ & $0.23\,\%$ & $0.38\,\%$ & $0.35\,\%$ & $0.43\,\%$ &
$1.34\,\%$ & $1.99\,\%$ & $2.29\,\%$\\
\bottomrule
\end{tabular}
\end{center}
\caption{Figure~\ref{fig:curvaturetest}b in numbers. A predicted departure
$\delta=|\mathcal{O}^\CD| D^2/2$ of the ratio $D_M'/D_H$ from the FLRW value $1$, third row. 
The $1/D^2$ weighting is what makes $\mathcal{O}^\CD$ large at low redshift: the underlying distance
effect is smallest there, and the requirement is accordingly hardest, $0.2$--$0.4\,\%$
below $z\simeq0.5$. At the Lyman-$\alpha$ point it relaxes to $2.3\,\%$, already below
the DESI DR2 per-tracer precision of $1.36\,\%$ on $D_M/r_d$ and $1.17\,\%$ on
$D_H/r_d$. What is missing is not the precision of the distances but the slope
$D_M'$, which no single tracer measures.}
\label{tableX}
\end{table*}

This is a sharper prediction than $\CK^\CD$ itself: 
$\mathcal{O}^\CD$ and $\CK^\CD$ coincide at $z=0$ and are within $50\,\%$ of each other
below $z\simeq0.5$; beyond $z\simeq1$, $|\mathcal{O}^\CD|$ is smaller than $|\CK^\CD|$
(at $z=1$: $0.015$ against $0.058$).
While $\CK^\CD$ vanishes at $z_\times$, $\mathcal{O}^\CD$ does not
($\mathcal{O}^\CD(z_\times)=+0.059$). $\CK^\CD$ therefore has a blind spot exactly where the
physics is most interesting, and $\mathcal{O}^\CD$ does not.
With regard to the significances in Table~\ref{tableVIII} and Table~\ref{tableIX}, $\mathcal{O}^\CD$ is the better measured at low redshift. The reason is that integration along the line of sight smoothes the
derivative that $\CK^\CD$ needs. The two routes (CPL and nonparametric) agree for $\mathcal{O}^\CD$ within $1\sigma$
everywhere ($\le0.93\sigma$), against $1.46\sigma$ for $\CK^\CD$.

A dark-energy fluid with the same $E(z)$ reduces the curvature diagnostic to the Friedmannian parameter, $\mathcal{O}^\CD = \Omega_k$, so this test directly discriminates the evolution of physical variables from the injected constant-curvature evolution in the standard model (unlike $\CK^\CD$ also for nonvanishing $k$), i.e. \textit{any} redshift dependence of ${\mathcal O}^\CD$ falsifies the whole FLRW class, independently of the dark energy equation of state.

Expressing \eqref{CBL} in terms of $D_M$, $r_d$ and $\kappa$, with $D=(D_M/r_d)/\kappa$ and $E=\kappa/(D_H/r_d)$, we write:
\begin{equation}
\label{CBL2}
\mathcal{O}^\CD = \frac{(E D')^2-1}{D^2} = \kappa^2\,\frac{(D_M'/D_H)^2-1}{(D_M/r_d)^2}\; ,
\end{equation}
so $\kappa$ cancels from the shape and sets only the amplitude: $\mathcal{O}^\CD$ is
the statement that $D_M'\ne D_H$, which is the FLRW identity. A predicted
$\mathcal{O}^\CD$ is therefore a predicted fractional offset
$\delta=|\mathcal{O}^\CD|D^2/2$ of $D_M'/D_H$ from unity, and this can be compared with a BAO error bar and sets a concrete observational program for our benchmark, 
\textit{cf.} Figure~\ref{fig:curvaturetest} and Table~\ref{tableX}.\footnote{A similar proposal and comparison with the curvature consistency function of \cite{tests} has been devised in \cite{wiegand2010} (\textit{cf.} Fig.2 therein), however, by using an evolving constant-curvature template metric instead of R\"as\"anen's propagation model.}

As Figure~\ref{fig:curvaturetest} shows, the requirement of discrimination is hardest at low redshift and only relaxes to the percent level beyond $z\simeq1.3$. Comparing with Figure~\ref{fig:koksbangbound}, the two tests therefore ask for different things: for $\CK^\CD$ it is the extension below $z=0.5$ that would strengthen the test of our benchmark, since the predicted signal grows by an order of magnitude and the parametrization-free significance with it.
For $\mathcal{O}^\CD$ the ordering is reversed: its large low-redshift values are due to the $1/D^2$ weighting rather
than being a larger effect. Measured against what can be achieved, the requirement is met
first at the high-redshift end: at the Lyman-$\alpha$ point the needed $2.3\,\%$
already exceeds the DESI DR2 per-tracer precision, and what is missing there is a precision on the 
slope $D_M'$ rather than on the distances.

\section{Light propagation and the scope of the closure}
\label{distances}

R\"as\"anen \cite{rasanen2009} develops an approximation for light propagation in  a statistically homogeneous and
isotropic dust universe with vanishing vorticity, also assuming that the null
shear contribution is negligible. The mean angular diameter distance $\bar D_A$ obeys
a Sachs equation whose only source is Ricci focusing of the matter content, i.e. 
$R_{\mu\nu}k^\mu k^\nu = 8\pi G\,T_{\mu\nu}k^\mu k^\nu$ (with the components of the 4-Ricci tensor $R_{\mu\nu}$, the 4-stress-energy tensor $T_{\mu\nu}$ and the light-propagation 4-vector $k^\mu$).
In the analyzed backreaction model the energy source is dust. Written in terms of redshift, 
\begin{align}
 H_\CD\,\frac{\mathrm{d}}{\mathrm{d}z}\!\left[(1+z)^2 H_\CD
 \frac{\mathrm{d}\bar D_A}{\mathrm{d}z}\right]=-4\pi G\av{\varrho}\,\bar D_A \; .
\label{eq:syksy}
\end{align}
In a FLRW model with a dark-energy fluid the same equation holds with $\av{\varrho}$
replaced by the homogeneous density $\varrho_m$ plus the sources of the postulated dark energy fluid, $\varrho_{\rm DE}+p_{\rm DE}=(1+w)\varrho_{\rm DE}$, 
\begin{equation}
 H\,\frac{\mathrm{d}}{\mathrm{d}z}\!\left[(1+z)^2H\frac{\mathrm{d}D_A}{\mathrm{d}z}\right]
 = -4\pi G\,(\varrho_m+\varrho_{\rm DE} + p_{\rm DE})\,D_A \; .
 \label{eq:friedmann}
\end{equation}
Subtracting the average dust density in the focusing formula \eqref{eq:syksy} from the effective focusing source and dividing by $H_\CD^2$, using \eqref{morphoncorrespondence}, we obtain:
\begin{align}
 \frac{4\pi G}{H_\CD^2}\Big[\big(\varrho^\CD_{\rm eff}+p^\CD_{\rm eff}\big)-\av{\varrho}\Big]
 =\frac{4\pi G\,\epsilon\dot\Phi^2_\CD}{H_\CD^2}\nonumber\\
 =-\frac{\CW_\CD+3\CQ_\CD}{6H_\CD^2}=\Omega^\CD_\CR+3\Omega^\CD_\CQ ={\mathcal K}^\CD\; .
 \label{eq:ident}
\end{align}
Thus, by using R\"as\"anen distances the missing focusing source is the morphon kinetic term \eqref{eq:kok}, i.e. the quantity analyzed in  Appendix~\ref{koksbang}.

The two focusing versions share the same expansion history and differ in their distances by the morphon kinetic term.
The mismatch vanishes iff $w^\Phi_\CD=-1$, equivalent to a cosmological constant
($\CW_\CD=-3\CQ_\CD$), and instantaneously at the phantom divide. 
For the observables we infer that $D_H=c/H_\CD$ measures the volume Hubble
functional directly and is not affected, no focusing equation enters it.
The transverse comoving distance $D_M$, luminosity distances, and the CMB acoustic
angle $\theta_*=r_s/D_M(z_*)$ are affected. Integrating the two different versions for
the baseline CPL closure gives the values in Table~\ref{tableXI}. 
The sign change near $z\simeq1.3$ is inherited from
the phantom divide and makes the low-redshift effect partly self-cancelling; the CMB,
which integrates through the whole negative tail, does not benefit from that
cancellation.

The reconstruction to obtain Table~\ref{tableXI} and Figure~\ref{fig:rasanen} is that of Appendix~\ref{nonparametric}, however,  with the data restricted to the six $D_H/r_d$ measurements and the Gaussian prior on $\kappa=c/(H_\DT r_d)$. The transverse points $D_M/r_d$ are dropped, and so is the BGS $D_V/r_d$, which mixes
$D_M$ and $D_H$; the six $E(z_i)$ are then independent and the within-tracer
correlations do not enter. ($K=3$ as before, $\chi^2=2.29$ for $3$ degrees of freedom
($6.91/10$ for the full set), with $4000$ Monte Carlo realizations; with no data below
$z=0.510$, the value at the pivot is an extrapolation of the smooth fit, which is why its
error grows only mildly).

Re-emphasizing the scope of the present closure, where we equate $w^\Phi_\CD$ to $w_{\rm DESI}$, 
the data combinations used previously are self-consistent at the level of the
radial BAO, while the transverse BAO, the supernova distances and the CMB angle are
systematically shifted: by less than $0.25\,\%$ for the galaxy and quasar tracers,
by $0.50\,\%$ at the Ly$\alpha$ point, and by $2.06\,\%$ for $\theta_*$, the last being the
only one that exceeds its measurement precision ($0.03\,\%$), (and it is comparable in magnitude, with opposite sign, to the second-order lensing shift of $1.4\,\%$ at $z^*$~\cite{genevaCMB}, \textit{cf.}~subsection~\ref{onlyapprox}.b). 
When using R\"as\"anen's distances, the main drawback of the approximation consists in following the fits all the way to
$z_*\simeq1100$; within the DESI range the induced shifts stay well inside the
statistical errors of the reconstruction. 
We therefore have to read the redshift-dependent statements that involve transverse data with care since they rely on the bias of Friedmannian transverse distances. Also, calibrations that involve the FLRW background and perturbations thereof, 
e.g.~the SNe pipeline, indicate that our non-FLRW approach does not directly profit from such data. Using BAO as a standard ruler is also questionable in the inhomogeneous reality \cite{boudBAO}. That said, it is nevertheless encouraging that our benchmark statements on the backreaction variables stay in place; $\Omega^\CD_\CQ, \Omega^\CD_\CW$ at $z\lesssim1$ are stable at the $\sim0.03$ level. 
This is also supported by Table~\ref{tableXI} and Figure~\ref{fig:rasanen}, a version of the closure that uses no 
transverse distances: the radial BAO fixes the expansion rate directly, so $H_\CD(z)$ follows without solving the focusing equation. What is important is that the result appears to leave the redshift dependence of the backreaction variables robust.

\begin{table*}
\begin{center}\small
\begin{tabular}{lccc|ccc}
\toprule
 & \multicolumn{3}{c|}{$\Omega^\CD_\CQ$} & \multicolumn{3}{c}{$\Omega^\CD_\CW$}\\
$z$ & radial only & all BAO & CPL & radial only & all BAO & CPL\\
\midrule
$0$     & $-0.196\pm0.066$ & $-0.249\pm0.038$ & $-0.214\pm0.029$ & $+0.876\pm0.066$ & $+0.931\pm0.038$ & $+0.895\pm0.029$\\
$0.294$ & $-0.217\pm0.014$ & $-0.221\pm0.012$ & $-0.242\pm0.008$ & $+0.755\pm0.011$ & $+0.747\pm0.009$ & $+0.766\pm0.012$\\
$0.5$   & $-0.215\pm0.017$ & $-0.201\pm0.012$ & $-0.220\pm0.010$ & $+0.649\pm0.030$ & $+0.623\pm0.018$ & $+0.635\pm0.016$\\
$1$     & $-0.175\pm0.029$ & $-0.148\pm0.015$ & $-0.136\pm0.008$ & $+0.393\pm0.030$ & $+0.373\pm0.020$ & $+0.349\pm0.012$\\
$1.5$   & $-0.097\pm0.017$ & $-0.086\pm0.011$ & $-0.075\pm0.004$ & $+0.182\pm0.025$ & $+0.192\pm0.022$ & $+0.181\pm0.011$\\
$2$     & $+0.006\pm0.033$ & $-0.016\pm0.028$ & $-0.041\pm0.002$ & $+0.035\pm0.044$ & $+0.073\pm0.032$ & $+0.097\pm0.009$\\
\bottomrule
\end{tabular}
\begin{tabular}{lccccccc}
\toprule
$z$ & $0.3$ & $0.5$ & $1.0$ & $1.5$ & $2.0$ & $2.33$ & $1089$\\
\midrule
$\bar D_A/D_A^{\rm FLRW}-1$ & $+0.153\,\%$ & $+0.230\,\%$ & $+0.145\,\%$ &
 $-0.107\,\%$ & $-0.358\,\%$ & $-0.501\,\%$ & $-2.06\,\%$\\
\bottomrule
\end{tabular}
\end{center}
\caption{Upper: the closure reconstructed from the DESI DR2 radial BAO alone  
against the full reconstruction of Appendix~\ref{nonparametric} and the CPL closure of the main text. No transverse
distance enters (i.e. no angular diameter or luminosity distance), so the construction is free of the focusing mismatch. At the pivot the shift on removing every $D_M$-type datum is $0.004$ in $\Omega^\CD_\CQ$
($0.3\sigma$) and $0.008$ in $\Omega^\CD_\CW$ ($0.7\sigma$). $\CQ_\CD(z_p)=1.93\pm0.08$ against
$1.92\pm0.07$. The qualitative structure is untouched: $\CQ_\CD>0$ with $\CQ_\CD (z=1) =$
($3.42\pm0.53$ radial-only, $2.92\pm0.28$ full), $\CW_\CD<0$ throughout. Hence, the closure barely moves.
Errors are $68\,\%$. $\chi^2=2.29$ for $3$ degrees of freedom. Lower: percentages as a function of redshift of the focusing mismatch of angular diameter distances with respect to FLRW-calibrated distances reaching percent level only if continued to last scattering.}
\label{tableXI}
\end{table*}

\begin{figure*}\centering
\includegraphics[width=0.8\textwidth]{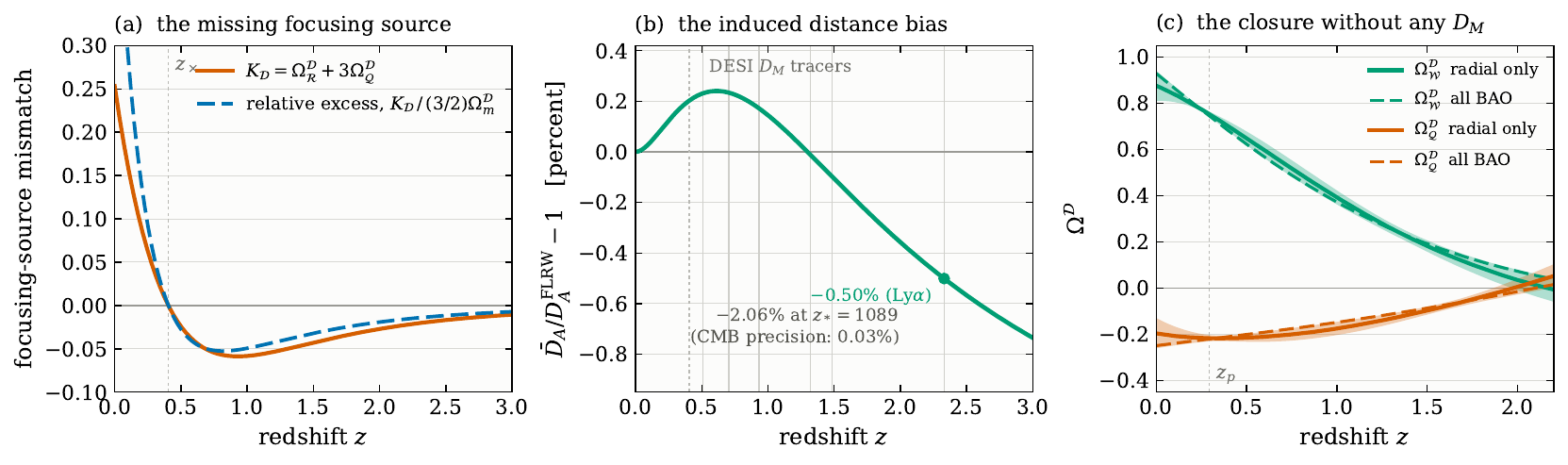}
\caption{Light propagation in the backreaction reading. (a) the mismatch between the
FLRW and the dust-only focusing sources, which by \eqref{eq:ident} is the morphon
kinetic term ${\mathcal K}_\CD$ of Appendix~\ref{koksbang} (solid), and the same relative to the dust source 
(dashed; clipped at $0.3$, $0.53$ at $z=0$); it changes sign at the phantom divide $z_\times=0.405$. (b) the
resulting bias of the mean angular diameter distance at fixed $H_\CD(z)$: the partial
cancellation across $z_\times$ keeps it below $0.5\,\%$ over the DESI BAO range
(vertical lines: the six tracers), while at last scattering it reaches $-2.06\,\%$. 
(c) the closure reconstructed from the radial BAO
alone (bands, $68\,\%$), which uses no transverse distance at all, against the full
reconstruction of Appendix~\ref{nonparametric} (dashed).}
\label{fig:rasanen}
\end{figure*}

\end{document}